\documentclass[prb,twocolumn,preprintnumbers,amsmath,amssymb,floatfix,showpacs]{revtex4}

\usepackage[dvips]{graphics,graphicx,color}

\newcommand{\vect}[1]{\boldsymbol{#1}}

\begin{document}

\title{Ground state properties of the one-dimensional electron liquid}

\author{R.\ M.\ Lee}

\affiliation{TCM Group, Cavendish Laboratory, University of Cambridge, J.\ J.\
Thomson Avenue, Cambridge CB3 0HE, United Kingdom}

\author{N.\ D.\ Drummond}

\affiliation{Department of Physics, Lancaster University, Lancaster LA1 4YB,
United Kingdom}

\begin{abstract}
  We present calculations of the energy, pair correlation function
  (PCF), static structure factor (SSF), and momentum density (MD) for
  the one-dimensional electron gas using the quantum Monte Carlo
  method.  We are able to resolve peaks in the SSF at
  even-integer-multiples of the Fermi wave vector, which grow as the
  coupling is increased. Our MD results show an increase in the
  effective Fermi wave vector as the interaction strength is raised in
  the paramagnetic harmonic wire; this appears to be a result of the
  vanishing difference between the wave functions of the paramagnetic
  and ferromagnetic systems. We have extracted the Luttinger liquid
  exponent from our MDs by fitting to data around $k_{\mathrm F}$,
  finding good agreement between the exponent of the ferromagnetic
  infinitely-thin wire and the ferromagnetic harmonic wire.
\end{abstract}

%We reduce finite-size effects in our calculation of the SSF by fitting a
%decaying oscillatory function to the tail of the PCF\@.

\pacs{71.10.Hf, 71.10.Pm, 73.63.Nm}

\maketitle

\section{Introduction \label{sec:introduction}}
Landau's theory of Fermi liquids has proven tremendously successful at
describing a wide range of systems of interacting fermions. In particular, the
theory legitimizes the free electron model by casting fermionic systems in
terms of weakly-interacting quasiparticles. Systems of electrons in 1D provide
an intriguing example of departure from the Landau Fermi liquid paradigm,
exhibiting non-Fermi-liquid behavior for any finite strength of the
electron-electron interaction.\cite{Giuliani_2005} Perhaps the simplest model
of electrons in one dimension is the one-dimensional (1D) homogeneous electron
gas (HEG), which comprises electrons on a uniform positively-charged
background.\cite{Giuliani_2005}

The strong correlation occurring in 1D ensures that the excitations are not
electron-like quasiparticles, but are instead collective in nature. An
appropriate description of the low-energy spectrum of the 1D HEG comes from
the theory of Tomonaga and
Luttinger.\cite{Tomonaga_1950,Luttinger_1963,Haldane_1981} There are several
experimental signatures of the Tomonaga-Luttinger (TL) liquid which
distinguish it from the normal Fermi liquid; these are largely accessible to
transport and tunneling experiments. For example, the conductivity of a 1D
channel as a function of the temperature is expected to vary logarithmically
in the presence of weak disorder for the Fermi liquid, and as a power law for
the TL liquid.\cite{Mitin_2009,Bockrath_1999} Analogous relations hold for the
differential conductivity and the optical conductivity. Also associated with
the lack of quasiparticles in the TL liquid is spin-charge separation, whereby
spin and charge excitations propagate at different characteristic
velocities.\cite{Haldane_1981,Voit_1994,Schulz_1998}

One-dimensional models are easy to envisage, but experimental observation of
1D behavior is potentially problematic. Low-dimensional systems are never
entirely independent of their 3D environment, leading to effects which have
the potential to obscure the 1D behavior. Furthermore, the presence of
impurities has been shown to alter drastically the behavior of a TL
liquid.\cite{Giuliani_2005,Meden_2002} However, even in manifestly 3D systems,
behavior unambiguously characteristic of electrons in 1D arises surprisingly
frequently. Features associated with the Luttinger model have been observed in
organic conductors (e.g., tetrathiafulvalene-tetracyanoquinodimethane and the
Bechgaard
salts),\cite{Ito_2005,Dressel_2005,Lorenz_2002,Schwartz_1998,Vescoli_2000,Dressel_2001}
transition metal oxides,\cite{Maekawa_2004,Hu_2002} carbon
nanotubes,\cite{Dresselhaus_2001,Egger_1998,Bockrath_1999,Ishii_2003,Shiraishi_2003}
edge states in quantum Hall
liquids,\cite{Milliken_1996,Chang_2003,Mandal_2001} semiconductor
heterostructures,\cite{Steinberg_2006,Zaitsev_2000,Liu_2005,Goni_1991,Auslaender_2000}
confined atomic gases,\cite{Moritz_2005,Recati_2003,Monien_1998} and atomic
nanowires.\cite{Schaefer_2008} Theoretical work on electrons in 1D thus has a
large region of potential applicability.

The exactly-solvable Luttinger model describes electrons moving in one
dimension with short-range interactions and linear dispersion. Studies
with long-range interactions have found that the exponents and
excitation velocities are nontrivially altered.\cite{Schulz_1993} One
thus expects to be able to describe the 1D HEG within the Luttinger
model framework, but the exact behavior of the parameters of the model
is largely unclear. The interactions that we study here are
long-ranged, possessing a $1/|x|$ Coulombic tail. This is most
applicable to systems where screening is a small effect, such as
isolated metallic carbon nanotubes and semiconductor structures where
there is negligible coupling to the substrate.

The 1D HEG has been studied with a variety of theoretical and computational
approaches. The principal distinction between various studies is the choice of
electron-electron interaction. The bare Coulomb interaction, $1/|x|$, which
describes an infinitely-thin wire, is perhaps conceptually the simplest
choice, although it is largely avoided in the
literature\cite{Astrakharchik_2010} in its original form due to the divergence
at $x=0$. Instead, many previous authors have removed the singularity while
retaining the long-range behavior by investigating interaction potentials of
the form $V(x)\propto (x^2+d^2)^{-1/2}$, where $d$ is a parameter related to
the width of the wire. This interaction has been studied
analytically\cite{Schulz_1993,Fogler_2005} and
numerically.\cite{Creffield_2001}

Otherwise, one can derive an effective 1D interaction by factorizing the wave
function into longitudinal and transverse parts and assuming that the
transverse component is the (2D) single-particle ground state of the confining
potential. The 1D interaction is then the matrix element of the 3D Coulomb
interaction with respect to the transverse
eigenfunctions.\cite{Fabrizio_1994,Casula_2006} An example of this is the
harmonic wire, in which the transverse confinement is provided by a parabolic
potential, leading to a Gaussian density profile in the transverse plane. The
harmonic wire has been studied with quantum Monte Carlo
(QMC),\cite{Casula_2006,Shulenburger_2008,Malatesta_2000} variants of the
Singwi-Tosi-Land-Sj\"olander
approach,\cite{Friesen_1980,Camels_1997,Tas_2003,Garg_2008} and the Fermi
hypernetted-chain approximation.\cite{Asgari_2007}

We have studied both the infinitely-thin wire and the harmonic wire using
QMC\@. In this article we report QMC calculations of the momentum density
(MD), energy, pair-correlation function (PCF), and static structure factor
(SSF) of the infinitely-thin wire at a variety of densities and system
sizes. The MD results in particular show the non-Fermi-liquid character of the
system and allow us to recover one of the TL parameters. The total
energy data that we provide are exact and may be regarded as a benchmark for
future work. We also present calculations of the MD for the
harmonic wire, again extracting one of the TL parameters.

The rest of this paper is structured as follows: the models for which we
perform our calculations are described in Sec.\ \ref{sec:models}. In Sec.\
\ref{QMC_details} we outline QMC methods and provide the details of our
approach. We report the ground state energies of both models in Sec.\
\ref{sec:energies} and describe the PCFs in Sec.\ \ref{sec:pcf}. In Sec.\
\ref{sec:ssf} we give the SSFs that we find for the infinitely-thin wire and
in Sec.\ \ref{sec:md} we give the MDs for both models. We describe the
procedure for estimating a parameter of the TL model in Sec.\
\ref{sec:ll_params}. Finally, we draw our conclusions in Sec.\
\ref{sec:conclusions}.  We use Hartree atomic units
($\hbar=|e|=m_e=4\pi\epsilon_0=1$) throughout this article.

\section{Models\label{sec:models}}
\subsection{Hamiltonian}
The Hamiltonians for both of the models we have studied may be written as
\begin{equation}
\hat{H}=-\frac{1}{2}\sum_{i=1}^N \frac{\partial^2}{\partial x_i^2}
+\sum_{i<j}V(x_{ij})+\frac{N}{2} V_{\rm Mad}\;,
\label{eq:H_inf_thin}
\end{equation}
where $V_{\rm Mad}$ is the Madelung energy (the interaction of a particle with
its own background and periodic images), $x_{ij}=|x_i-x_j|$ is the distance
between electron $i$ and electron $j$, and $V(x_{ij})$ is the Ewald
interaction; this is the interaction of an electron at $x_i$ with another
electron at $x_j$, all of electron $j$'s periodic images, and $1/N$-th of the
uniform positive background. The two models that we have studied differ in the
$V(x_{ij})$ and $V_{\rm Mad}$ terms.

\subsection{Infinitely-thin wire}
The Ewald interaction for the infinitely-thin wire may be written
\begin{eqnarray}
V(x_{ij}) & = & \sum_{n=-\infty}^{\infty}  \left (
\frac{1}{|x_{ij}+nL|}\right. \nonumber \\ & & \hspace{3em} \left. {}
-\frac{1}{L}\int^{L/2}_{-L/2} \frac{dy}{|x_{ij}+nL-y|}\right ),
\label{eq:inf_thin_int}
\end{eqnarray}
which is calculated in practice using an accurate approximation based on the
Euler-Maclaurin summation formula; see Eq.\ (4.8) of Ref.\
\onlinecite{Saunders_1994} for details.

The interaction of Eq.\ (\ref{eq:inf_thin_int}) diverges as $1/x_{ij}$
when {$x_{ij}\rightarrow 0$}. In higher dimensions, the divergence in
the interaction energy is canceled by an equal and opposite divergence
in the kinetic energy, so that nodes do not necessarily occur where
two antiparallel spins occupy the same position.\cite{Kato_1957} In
the infinitely-thin 1D system, the curvature of the wave function is
unable to compensate for the divergence in the interaction potential,
so the trial wave function has nodes at all of the coalescence points
for both parallel and antiparallel spin pairs. The result is that
the ground state energy is independent of the spin-polarization and
depends only on the density. In other words, the Lieb-Mattis
theorem\cite{Lieb_1962} does not apply and the paramagnetic and
ferromagnetic states are degenerate for the interaction of Eq.\
(\ref{eq:inf_thin_int}). We have examined only the fully
spin-polarized case for the infinitely-thin wire.

\subsection{Harmonic wire}
The second model we have studied describes electrons in a 2D confinement
potential given by
\begin{equation}
V_{\perp}(r_\perp)=r^2_{\perp}/8b^4,
\label{eq:transpot}
\end{equation}
where $b$ is the width parameter and $r$ is the magnitude of the projection of
the electron position onto the plane perpendicular to the axis of the
wire. The Ewald-like interaction for this model may be written
as\cite{Casula_2006,Malatesta_thesis}
\begin{eqnarray}
V(x_{ij})&=& \sum_{m=-\infty}^{\infty}\Bigg \lbrace \frac{\sqrt{\pi}}{2b}
e^{(x_{ij}-mL)^2/(4b^2)}{\rm erfc} \left (
\frac{|x_{ij}-mL|}{2b}\right )\nonumber \\ &-&\frac{1}{|x_{ij}-mL|} \; {\rm
erf} \left (  \frac{|x_{ij}-mL|}{2b}\right ) \Bigg \rbrace\nonumber \\
&+&\frac{2}{L}\sum_{n=1}^\infty E_1 \left [(bGn)^2 \right ]\cos(Gnx_{ij}) \;,
\label{eq:Ewald_like_sum}
\end{eqnarray}
where $G=2\pi/L$. Equation (\ref{eq:Ewald_like_sum}) possesses a
long-range Coulomb tail and is finite at $x_{ij}=0$. A derivation of
Eq.\ (\ref{eq:Ewald_like_sum}) is given in Appendix
\ref{app:ewald}. For the harmonic wire we have probed different
polarizations, $\zeta=|N_\uparrow-N_\downarrow|/N$.

\section{Details of calculations\label{QMC_details}}
We computed expectation values using the variational and diffusion Monte Carlo
(VMC and DMC, respectively) methods as implemented in the \textsc{casino}
program.\cite{casino2} For the infinitely-thin wire, we combined VMC and DMC
results to form extrapolated estimates\cite{Foulkes_2001} where applicable,
whereas for the harmonic wire we used VMC alone.

In the VMC method the expectation value of the Hamiltonian with respect to a
trial wave function is calculated using a stochastic integration
technique.\cite{Foulkes_2001} Trial wave functions usually contain a number of
free parameters; we optimized the free parameters in our wave function by
unreweighted variance minimization\cite{Umrigar_1988a,Kent_1999,Ndd_newopt}
and linear-least-squares energy minimization.\cite{Umrigar_emin} DMC is a
stochastic projector technique for solving the many-body Schr\"odinger
equation and generates configurations distributed according to the product of
the trial wave function and its ground state
component.\cite{Foulkes_2001,Ceperley_1980} DMC calculations of expectation
values of operators that commute with the Hamiltonian are in principle exact
for systems in which the wave function nodes are known; this is the case for
both the infinitely-thin and harmonic wires.

We used a Slater-Jastrow wave function for both systems, where the
Jastrow factor comprised two-body terms consisting of smoothly
truncated polynomials and a sum of cosines with periodicity
commensurate with that of the simulation cell.\cite{ndd_jastrow} The
orbitals in the Slater determinant were plane waves with wave vectors
up to {$k_{\mathrm F}=\pi/(4r_s)$} for the paramagnetic systems and
{$k_{\mathrm F}=\pi/(2r_s)$} for the ferromagnetic systems. The
orbitals were evaluated at quasiparticle coordinates related to the
actual coordinates by a backflow transformation.\cite{Pablo_2006}
Backflow provides an efficient way of describing three-body
correlations in the 1D HEG, but leaves the exact nodal surface
unchanged.

%begin{table}
% \begin{center}
%  \vspace{0.5cm}
%  \begin{tabular}{l@{\extracolsep{20pt}} l}
%    \hline\hline
%    Method & \% $E_{\rm correlation}$ \\ \hline
%    DMC             &100\\
%    VMC SJ(2+3),BF  &99.9981(9)\\
%    VMC SJ(2+3)     &99.9979(9)\\
%    VMC SJ2,BF      &99.9989(9)\\
%    VMC SJ2         &99.9981(9)\\
%    HF              &0\\
%    \hline
%  \end{tabular}
%  \caption{The percentage of the correlation energy retrieved for
%    $r_s=15$ a.u.\ and $N=15$ with several wave function types. The
%    label VMC SJ(2+3),BF refers to the wave function with both two
%    and three-body terms in the Jastrow factor and the inclusion of
%    backflow transformations.  The error bars on the VMC energy were
%    $O(10^{-8})$ a.u. The numbers are similar for other
%    densities and system sizes.\label{table:backflow}}
%\end{center}
%\end{table}

One method for assessing the wave function quality is to examine the fraction
of the correlation energy retrieved, $(E_{\rm HF}-E_{\rm VMC})/(E_{\rm
HF}-E_{\rm DMC})$, where $E_{\rm HF}$ is the Hartree-Fock energy, and $E_{\rm
DMC}$ and $E_{\rm VMC}$ are the DMC and VMC energies, respectively. We tested
several types of wave function for the infinitely-thin wire with $r_s=15$
a.u., $N=15$, and $\zeta=1$; our VMC calculations retrieved $99.9989(9)\%$ of
the correlation energy when we used a two-body Jastrow factor and backflow
transformations [the error bars were $O(10^{-8})$ a.u.], which is the type of
wave function we use throughout this paper. While it is indeed the case that
DMC is formally exact for the 1D HEG, the quality of the trial wave function
is important for the statistical efficiency of the DMC method and the accuracy
with which expectation values of operators that do not commute with the
Hamiltonian may be computed.

\begin{center}
  \begin{figure}
    \includegraphics[scale=0.31]{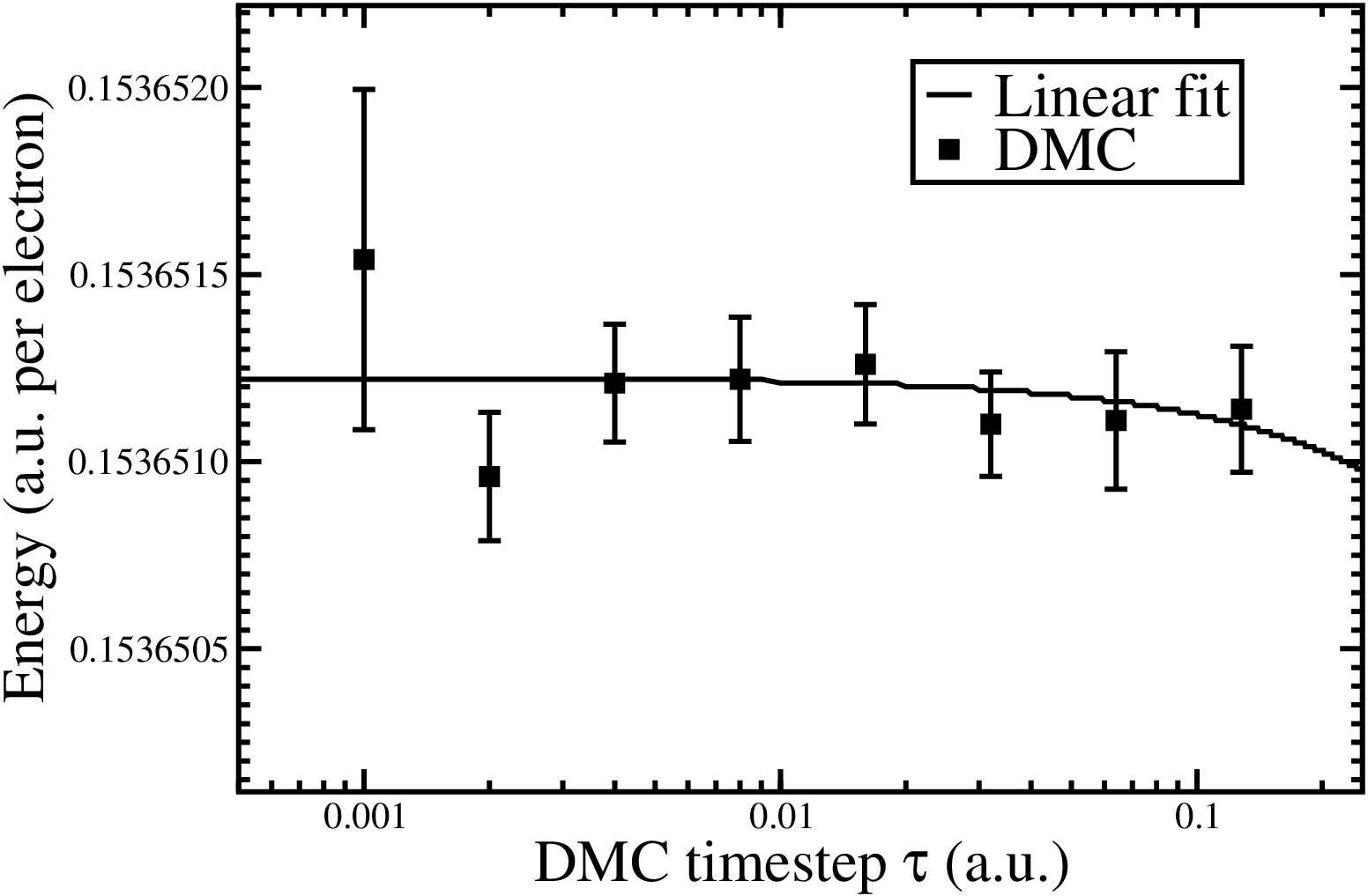}
    \caption{DMC energy of the infinitely-thin wire for several different
      timesteps and a linear fit to the data. The plot is for $r_s=1$ a.u.\
      and $N=37$ with 1000 configurations.\label{fig:dtdmc}}
  \end{figure}
\end{center}
The DMC energy did not change beyond statistical error upon varying the number
of walkers between 640 and 2000, so we used $\sim 1000$ walkers in our
calculations and assumed that population control bias is negligible. The
dependence of the energy upon the DMC timestep $\tau$ was also investigated;
Fig.\ \ref{fig:dtdmc} shows that for small $\tau$ the energy is constant. We
performed our calculations at a single timestep given by
$\tau=0.008\;\!r_s^2$. This fairly conservative choice was made to ensure that
time step bias is entirely negligible. The RMS distance diffused by each
electron in a single step was thus slightly less than $r_s/10$.

For the infinitely-thin wire, we used simulation cells containing 37, 55, 73,
and 99 particles subject to periodic boundary conditions for our calculations
of the energy, PCF, and SSF\@. Our MD calculations for the infinitely-thin
wire also used a much larger cell with $N=255$. For the harmonic wire, we used
cells with $N=123$, $155$, and occasionally $255$ for the $\zeta=1$ systems
and cells with $N=22$ and $102$ for the $\zeta=0$ systems.

Previous work encountered difficulties in sampling different spin
configurations of the harmonic wire for $\zeta \neq 1$ due to the presence of
``pseudonodes'' at the antiparallel coalescence points,\cite{Malatesta_2000}
although these problems were largely overcome by the use of
lattice-regularized diffusion Monte Carlo (LRDMC) in Ref.\
\onlinecite{Casula_2006}. The problem occurred because for strong, repulsive
interactions the wave function can become small when two antiparallel spins
approach one another. Combined with a small time step this can lead to
simulations where opposite spins exchange positions infrequently and the space
of spin configurations is explored very inefficiently. Use of a small time
step is a necessary part of the algorithm of projector methods like DMC\@. We
have avoided ergodicity problems by using VMC to study the harmonic wire; in
the VMC method there is no restriction other than ergodicity on the transition
probability density and one may propose moves however one wishes provided that
the acceptance probability is modified accordingly. We use
electron-by-electron sampling with the transition probability density given by
a Gaussian centered on the initial electron position. The VMC ``time step'' in
fact bears no relation to real time and is simply the variance of the
transition probability density.  In practice, the unmodified time steps
(chosen to achieve a $50\%$ acceptance ratio) used in VMC are usually large
enough to eliminate ergodicity problems, although we found some cases where it
was necessary to enforce a lower limit on the width of the transition
probability density. Table \ref{table:spinexchange} shows the frequency with
which electrons changed positions in our simulations for both high and low
density systems with strong and weak confinement.
\begin{table}
  \begin{center}
   \vspace{0.5cm}
   \begin{tabular}{l@{\extracolsep{20pt}} ll}
     \hline\hline

     $r_s$ (a.u.) & $b$ (a.u.) & $s_{\rm exch}$\\ \hline

     1 & 0.1 & 0.051(1)\\

     1 & 4   & 0.160(2)\\

     15 & 0.1 & 0.0016(2)\\

     15 & 4   & 0.0020(3)\\

     \hline \hline
   \end{tabular}
   \caption{Frequency with which electrons' paths cross in our VMC simulations
     of the harmonic wire. The quantity $s_{\rm exch}$ is the proportion of
     proposed single-electron moves that result in a change in the ordering of
     the particles. A typical calculation comprises between $10^7$ and $10^8$
     proposed single-electron moves. The data shown are for $N=22$.
     \label{table:spinexchange}}
 \end{center}
 \end{table}

\section{Results}
\subsection{Energies\label{sec:energies}}
\begin{table}
\begin{center}
\begin{tabular}{c@{~~~}r@{~~~~~~}r@{.}l}
\hline \hline

$r_s$ (a.u.) &$N$&
\multicolumn{2}{c}{$E_{\rm DMC}$ (a.u.\ / elec.)} \\

\hline

1&37& $0$&$1536513(3)$  \\

1&55& $0$&$1539427(2)$  \\

1&73& $0$&$1540497(3)$  \\

1&99& $0$&$1541147(2)$   \\

2&37& $-0$&$20637509(9)$   \\

2&55& $-0$&$20628042(7)$   \\

2&73& $-0$&$20624573(6)$   \\

2&99& $-0$&$20622457(9)$   \\

5&37& $-0$&$20397386(3)$   \\

5&55& $-0$&$20395138(2)$   \\

5&73& $-0$&$20394308(2)$   \\

5&99& $-0$&$20393799(2)$   \\

10&37& $-0$&$14288342(1)$   \\

10&55& $-0$&$14287568(1)$   \\

10&73& $-0$&$14287284(1)$   \\

10&99& $-0$&$142871058(9)$   \\

15&37& $-0$&$110474492(5)$   \\

15&55& $-0$&$110470307(4)$  \\

15&73& $-0$&$110468755(4)$   \\

15&99& $-0$&$110467811(5)$   \\

20&37& $-0$&$090782764(5)$   \\

20&55& $-0$&$090780068(2)$   \\

20&73& $-0$&$090779064(2)$   \\

20&99& $-0$&$090778454(2)$   \\

\hline \hline
\end{tabular}
\caption{The DMC energies for the infinitely-thin wire.
  \label{table:energies_infthin}}
\end{center}
\end{table}
\begin{center}
  \begin{figure}
    \includegraphics[scale=0.31]{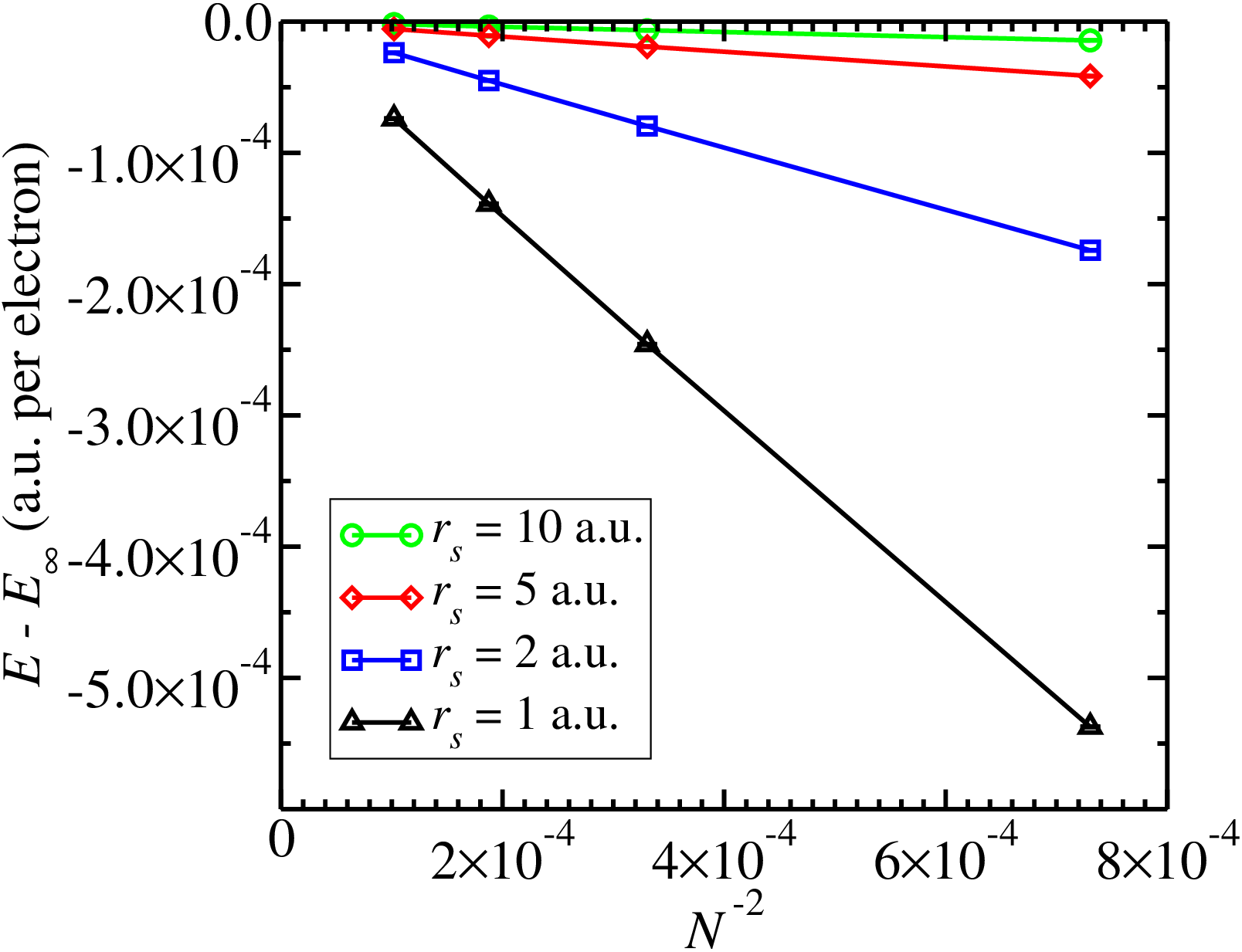}
    \caption{(Color online) Plot of the DMC energy against the
      reciprocal of the square of the system size for the infinitely-thin
      wire. The energy has been offset by the extrapolate $E_\infty$ that one
      obtains using the form $E(N)=E_{\infty}+BN^{-2}$.
      \label{fig:energy_extrap}}
  \end{figure}
\end{center}
\begin{table}
\begin{center}
\begin{tabular}{c@{~~~~}r@{.}l}
  \hline \hline

  $r_s$ (a.u.) &
  \multicolumn{2}{c}{$E_{\infty}$ (a.u.\ / elec.)} \\

  \hline

  1&0&1541886(2)\\

  2&$-0$&20620084(7)\\

  5&$-0$&20393235(2)\\

  10&$-0$&142869097(9)\\

  15&$-0$&110466761(4)\\

  20&$-0$&090777768(2)\\

  \hline \hline
\end{tabular}
\caption{The DMC energies for the infinitely-thin wire
  extrapolated to the thermodynamic limit using the form
  $E(N)=E_{\infty}+BN^{-2}$. \label{table:extrapolated_energies_infthin}}
\end{center}
\end{table}
% Our calculations with $k_s\neq 0$ confirm that this is the case.

For the infinitely-thin wire, we used DMC to calculate the exact ground state
energy since there is no ergodicity problem. Table
\ref{table:energies_infthin} shows the DMC energies that we obtained for
$r_s=1$, 2, 5, 10, 15, and 20 a.u., with $N=37$, 55, 73, and 99 particles. We
find that the Fourier transform of the two-body Jastrow factor $u(k)$ takes
the form $u(k)\propto 1/k$ as $k\rightarrow 0$, allowing us to estimate the
leading-order scaling of the finite-size correction to the kinetic
energy.\cite{Chiesa_2006} Furthermore, we observe that the static structure
factor $S(k)$ goes to zero linearly at $k=0$, allowing calculation of the
corresponding correction for the potential energy.\cite{Chiesa_2006} Motivated
by these results, we use the form $E(N)=E_{\infty}+BN^{-2}$ to extrapolate to
the thermodynamic limit. Figure \ref{fig:energy_extrap} demonstrates that this
form fits the data well, and Table \ref{table:extrapolated_energies_infthin}
shows the extrapolated energies $E_\infty$.

The many-body Bloch theorem states that the wave function $\psi_T$
satisfies\cite{Rajagopal_1995}
\begin{equation}
  \label{eq:manybodyBloch}
  \Psi_T(x_1,\ldots,x_j+L,\ldots,x_N)=e^{ik_sL}
\Psi_T(x_1,\ldots,x_N)\;,
\end{equation}
where $L$ is the length of the simulation cell and $k_s$ is the simulation
cell Bloch wave number. Averaging over $k_s$ in the irreducible Brillouin zone
(``twist averaging'') has been shown to reduce greatly single-particle
finite-size effects in two and three
dimensions.\cite{Ndd_2009,Holzmann_2009,Twist_averaging} In 1D, however, use
of a nonzero $k_s$ does not result in reoccupation of the orbitals, merely
adding $k_s^2/2$ to the energy per particle and leaving unchanged (or
trivially altering) other expectation values.  It is easy to show that the
average of $k_s^2/2$ over the Brillouin zone falls off as $O(N^{-2})$ in 1D;
hence single-particle finite-size effects simply lead to an additional
$O(N^{-2})$ error in the energy per particle, which is removed when we
extrapolate to infinite system size.

The trial wave functions in our calculations for the harmonic wire are of
sufficient quality that the variational energies we obtain are in statistical
agreement with exact results in the literature;\cite{Casula_2006} Table
\ref{table:lrdmc_comparison} shows the comparison.
\begin{table}
\begin{center}
\begin{tabular}{c@{~~~}r@{.}lr@{.}l}
\hline \hline

$r_s$ (a.u.) & \multicolumn{2}{c}{$E_{\rm VMC}$ (a.u.\ / elec.)\;\;} &
\multicolumn{2}{c}{$E_{\rm LRDMC}$ (a.u.\ / elec.)} \\

\hline

1     &  0&0901489(7)      &  0&09014(1)    \\

2     &  $-0$&1631207(8)     &  $-0$&16311(2)    \\

10    &  $-0$&1231560(3)     &  $-0$&123157(3)    \\

15    &  $-0$&0971194(1)     &  $-0$&097120(2)    \\

20    &  $-0$&0807160(2)     &  $-0$&080717(1)    \\

\hline \hline
\end{tabular}
\caption{Comparison of our VMC energies for the harmonic wire ($b=1$ a.u.,
  $\zeta=1$) with those of Ref.\ \onlinecite{Casula_2006}, calculated using
  the LRDMC method. For both sets of results the energies were extrapolated to
  the thermodynamic limit using the functional form
  $E(N)=E_{\infty}+BN^{-1}+CN^{-2}$, where $E_{\infty}$, $B$, and $C$ are
  fitting parameters.
  \label{table:lrdmc_comparison}}
\end{center}
\end{table}

\subsection{Pair-correlation function\label{sec:pcf}}
\begin{center}
  \begin{figure}
    \includegraphics[scale=0.31]{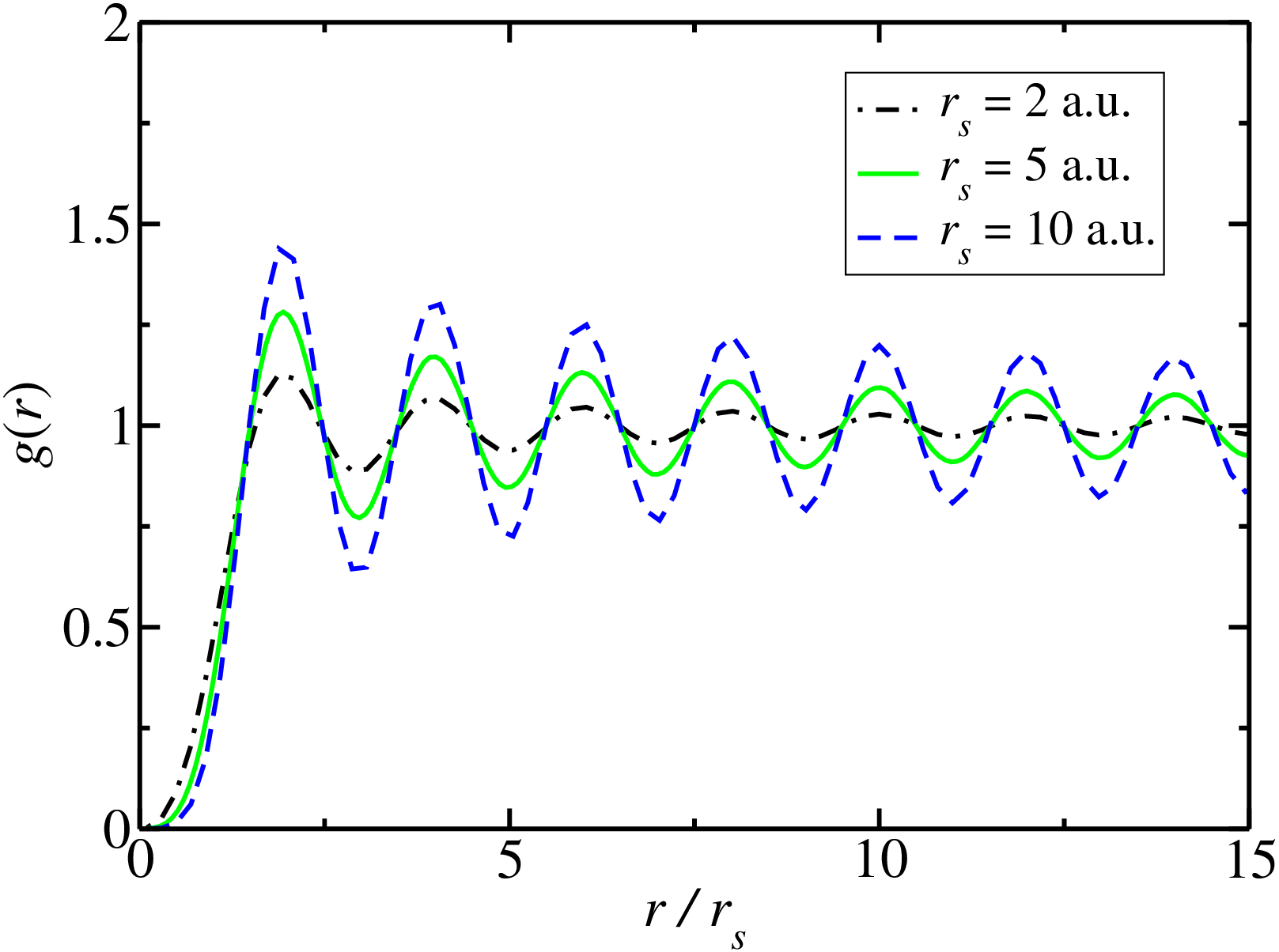}
    \caption{(Color online) PCF of the infinitely-thin wire at several
      densities. The data shown are for $N=99$ and are extrapolated estimates
      ${[2g_{\rm DMC}(x)-g_{\rm VMC}(x)]}$.
      \label{fig:pcf_plot}}
  \end{figure}
\end{center}
\begin{center}
  \begin{figure}
    \includegraphics[scale=0.31]{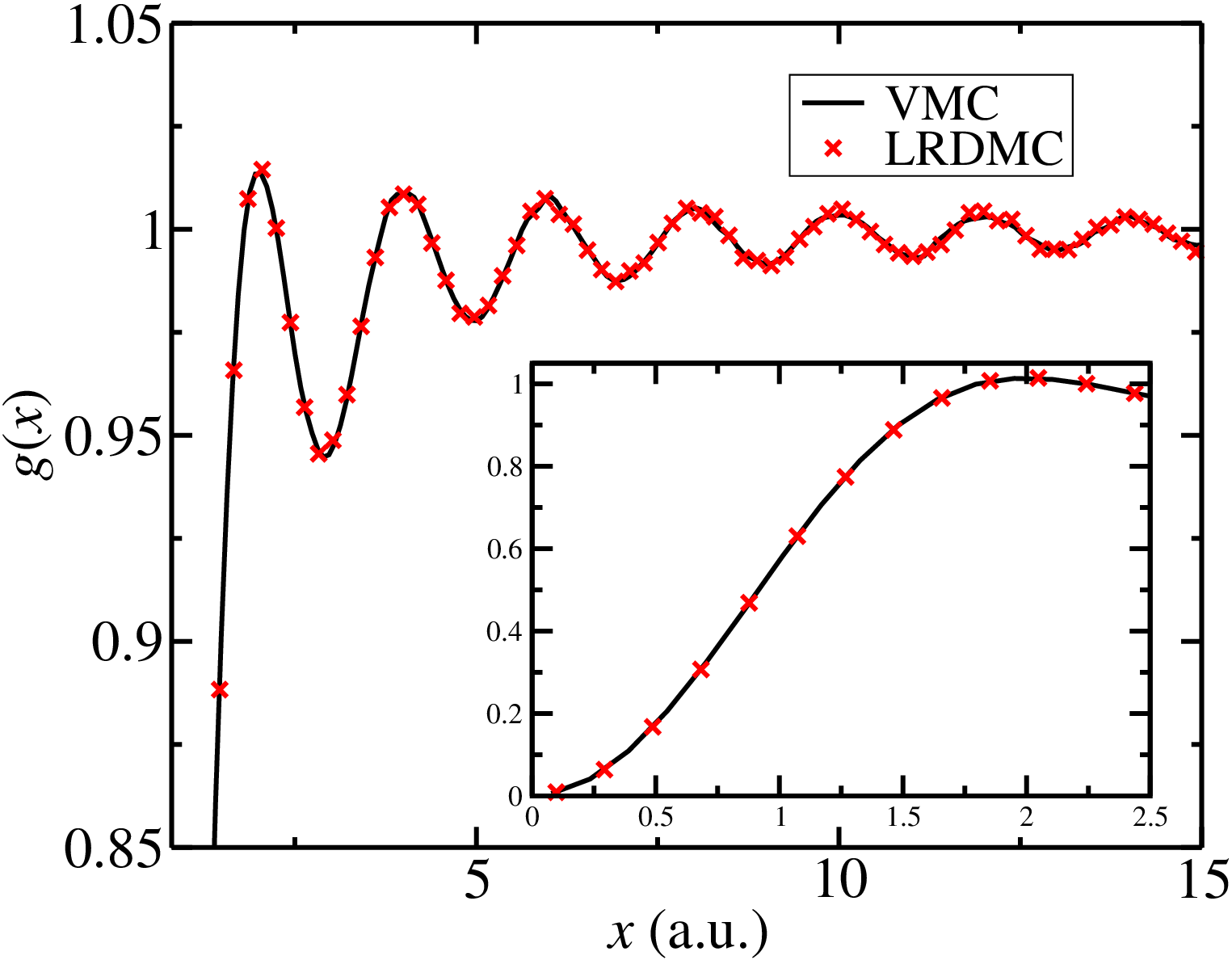}
    \caption{(Color online) PCF of the harmonic wire with $r_s=1$ a.u.,
      $N=39$, $b=1$ a.u., and $\zeta=1$. The solid line shows our VMC results
      and the symbols show the LRDMC results of Ref.\
      \onlinecite{Casula_2006}. The inset shows the same data at the
      origin. \label{fig:pcf_plot_casula}}
  \end{figure}
\end{center}
\begin{center}
  \begin{figure}
    \includegraphics[scale=0.31]{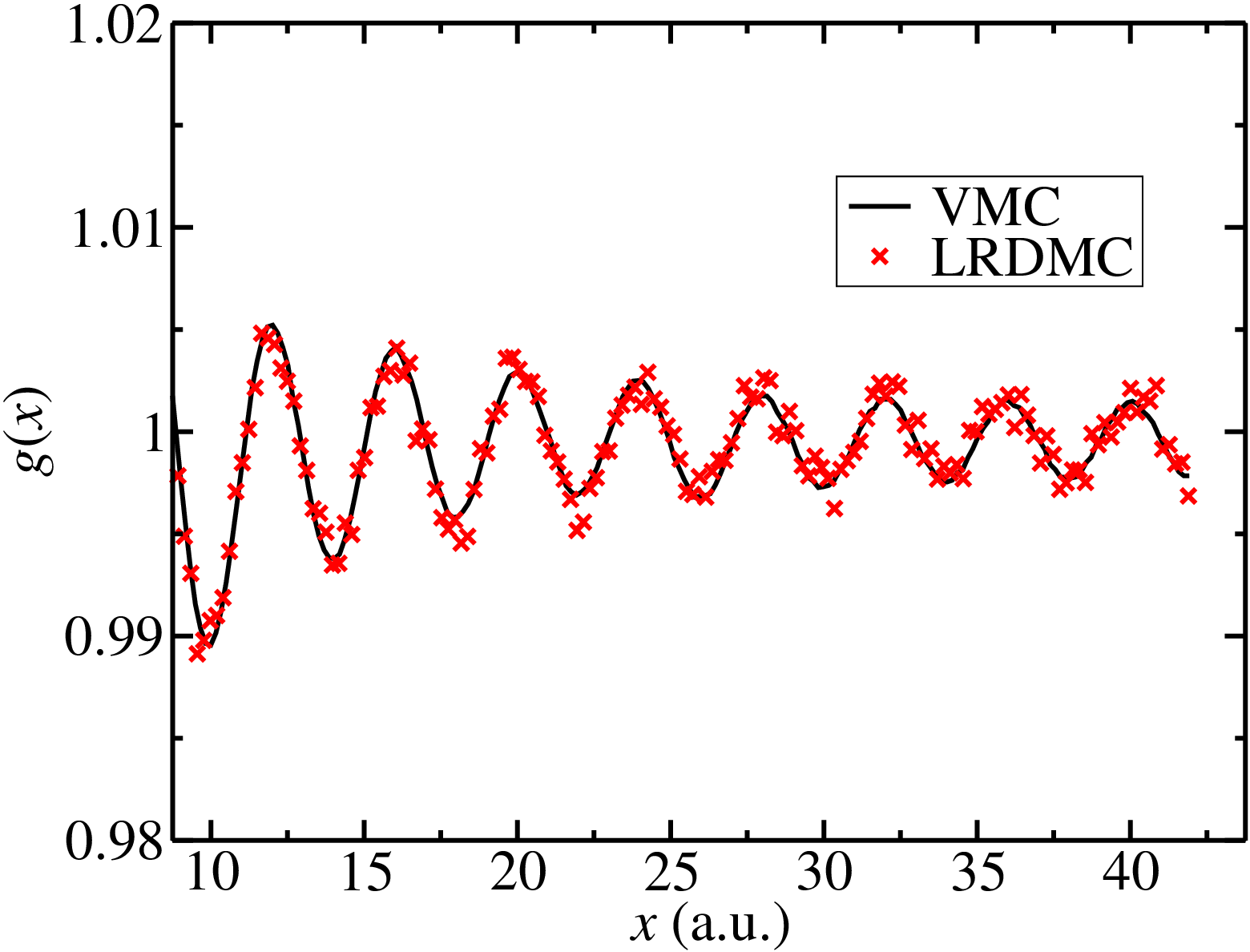}
    \caption{(Color online) PCF of the harmonic wire with $r_s=1$ a.u.,
      $N=42$, $b=1$ a.u., and $\zeta=0$. The solid line shows our VMC results
      and the symbols show the LRDMC results of Ref.\
      \onlinecite{Casula_2006}. The function plotted is
      ${[g_{\uparrow\uparrow}(x)+g_{\uparrow\downarrow}(x)]/2}$.
      \label{fig:pcf_plot_casula2}}
  \end{figure}
\end{center}
The PCF is accumulated in QMC simply by binning the interparticle distances
throughout the simulation.  The parallel-spin PCF is
\begin{equation}
  g_{\uparrow\uparrow}(x)=\frac{1}{\rho_{\uparrow}^2} \left \langle
  \sum_{i>j}^{N_{\uparrow}} \delta(|x_{i,\uparrow}-x_{j,\uparrow}|-x) \right
  \rangle\;,
\label{eq:pcf_parallel}
\end{equation}
where $\rho_\sigma$ is the average density of electrons with spin $\sigma$,
$x_{i,\sigma}$ is the position of the $i$th electron with spin $\sigma$ and
the angular brackets denote an average over the configurations generated by the
QMC algorithms. The antiparallel-spin PCF may be written as
\begin{equation}
  g_{\uparrow\downarrow}(x)=\frac{1}{\rho_{\uparrow}\rho_{\downarrow}} \left
  \langle  \sum_{i}^{N_{\uparrow}} \sum_{j}^{N_{\downarrow}}
  \delta(|x_{i,\uparrow}-x_{j,\downarrow}|-x) \right \rangle\;.
  \label{eq:pcf_antiparallel}
\end{equation}
The PCF for the harmonic wire was calculated for different confinements and
system sizes by Casula \textit{et al.}\ using the lattice-regularized DMC
method.\cite{Casula_2006} Figures \ref{fig:pcf_plot_casula} and
\ref{fig:pcf_plot_casula2} show the agreement of the LRDMC results with the
present work. Figure \ref{fig:pcf_plot} shows the PCF for the infinitely-thin
wire at several values of $r_s$.

\subsection{Static structure factor\label{sec:ssf}}

%\begin{center}
%  \begin{figure}
%    \includegraphics[scale=0.31]{rs1b1z0N42_ssf.eps}
%    \caption{(Color online) The static structure factor of the
%      harmonic wire with $r_s=1$ a.u., $N=42$, $b=1$ a.u., and $\zeta=0$. The
%      solid line shows our VMC data and the symbols show the LRDMC
%      results of Ref.\ \onlinecite{Casula_2006}. The PCF was not
%      extended beyond $L/2$ for this plot.\label{fig:casula_ssf_plot}}
%  \end{figure}
%\end{center}

\begin{center}
  \begin{figure}
    \includegraphics[scale=0.31]{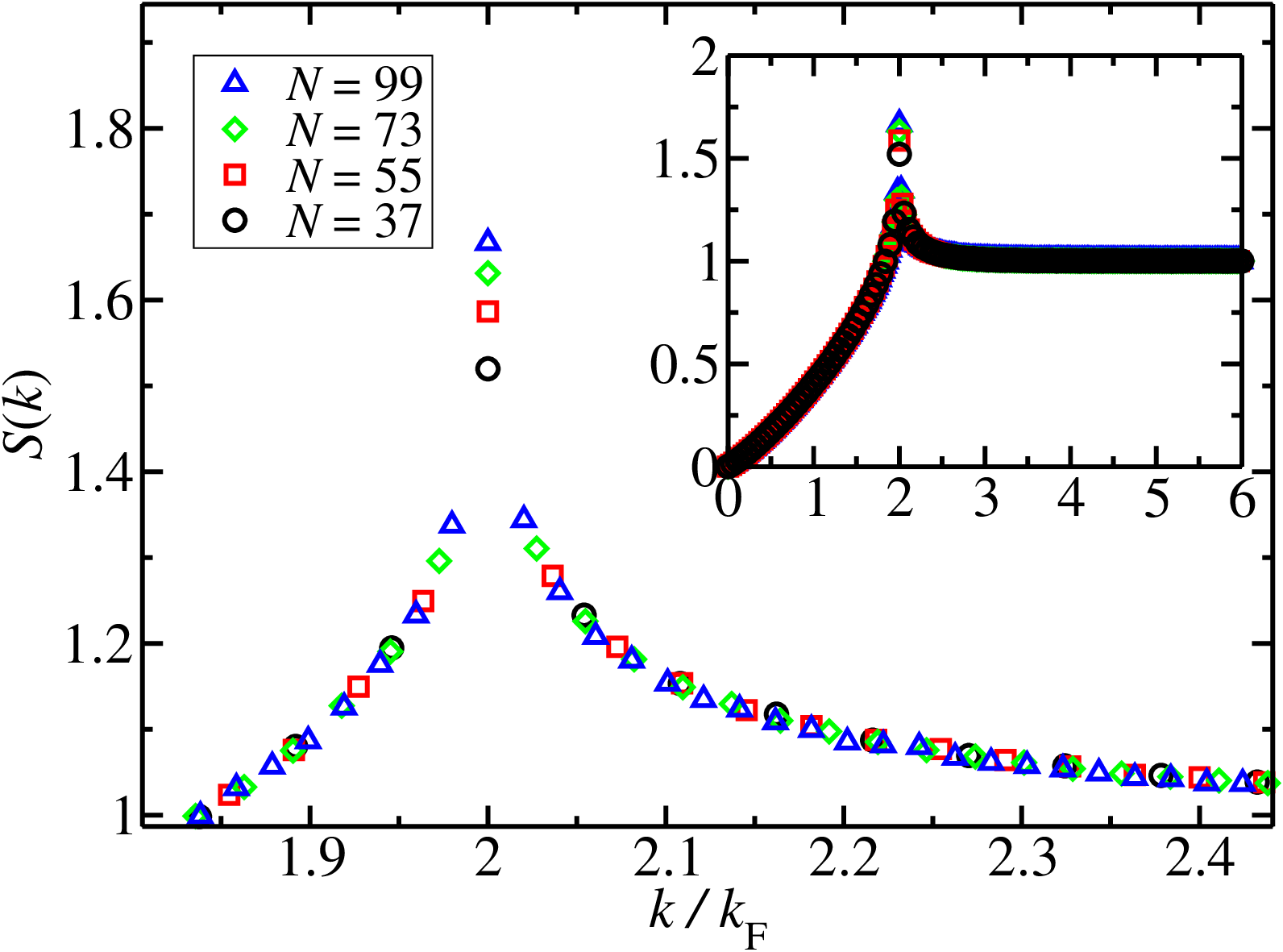}
    \caption{(Color online) SSF of the infinitely-thin wire at several system
      sizes. The data shown are extrapolated estimates ${[2S_{\rm
      DMC}(k)-S_{\rm VMC}(k)]}$ for $r_s=2\;{\rm a.u.}$. The main plot shows
      the behavior at the peak and the inset shows a zoomed-out view. The PCF
      was not extended beyond $L/2$.\label{fig:rs02_ssf_plot}}
  \end{figure}
\end{center}
\begin{center}
  \begin{figure}
    \includegraphics[scale=0.31]{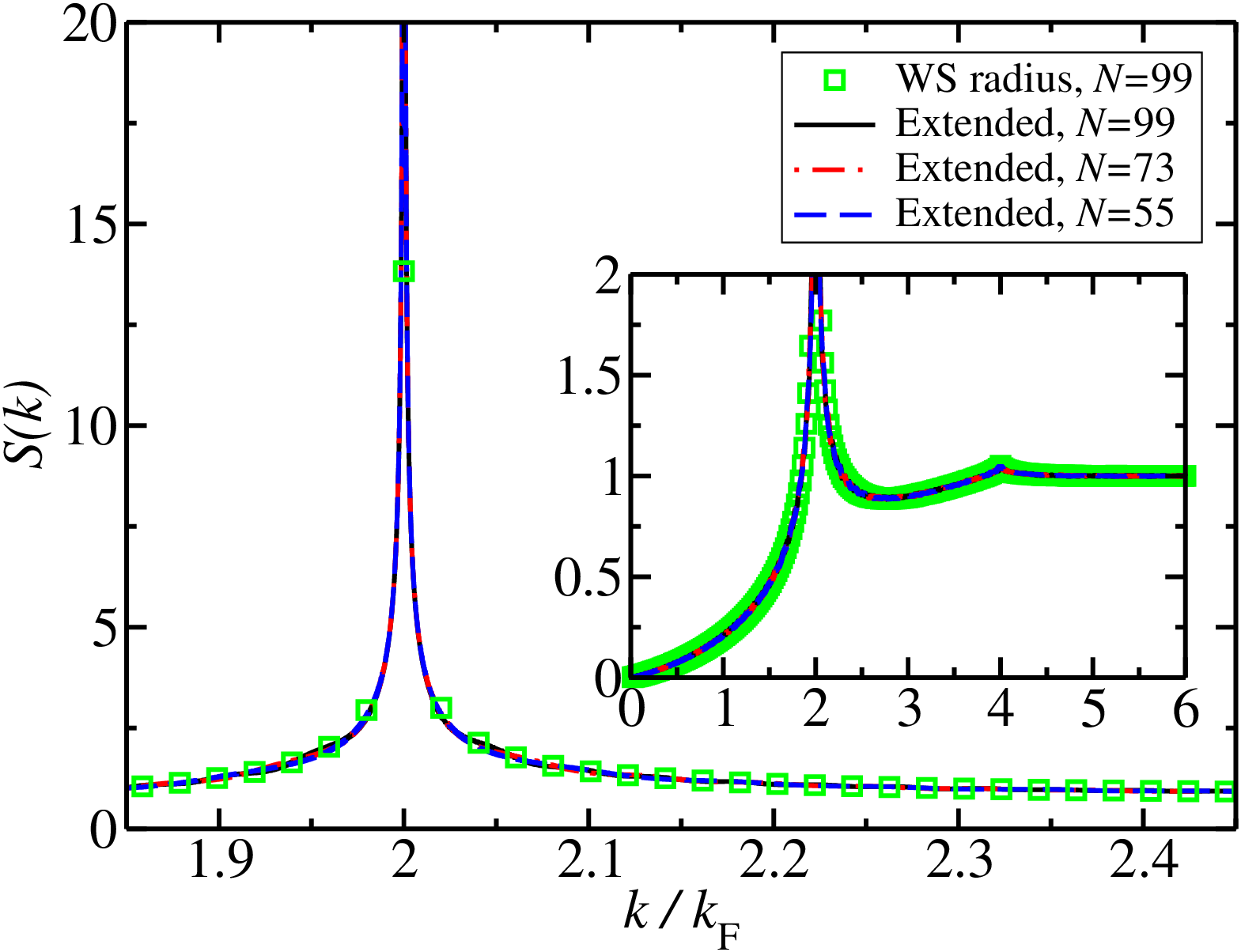}
    \caption{(Color online) Effect of extending the PCF before performing the
      transformation of Eq.\ (\ref{eq:ssf}). The square symbols (labeled ``WS
      radius'') show the SSF obtained from the finite-cell PCF\@. The solid,
      dash-dot, and dashed lines (which all lie on top of one another) are
      from the $N=99$, $73$, and $55$ PCFs, respectively, where in each case
      the PCF has been extended out to many simulation cell lengths using the
      fitting form of Eq.\ (\ref{eq:pcf_tails}). The data shown are for
      $r_s=20$ a.u.
      \label{fig:rs20_pcf_extended.eps}}
  \end{figure}
\end{center}
\begin{center}
  \begin{figure}
    \includegraphics[scale=0.31]{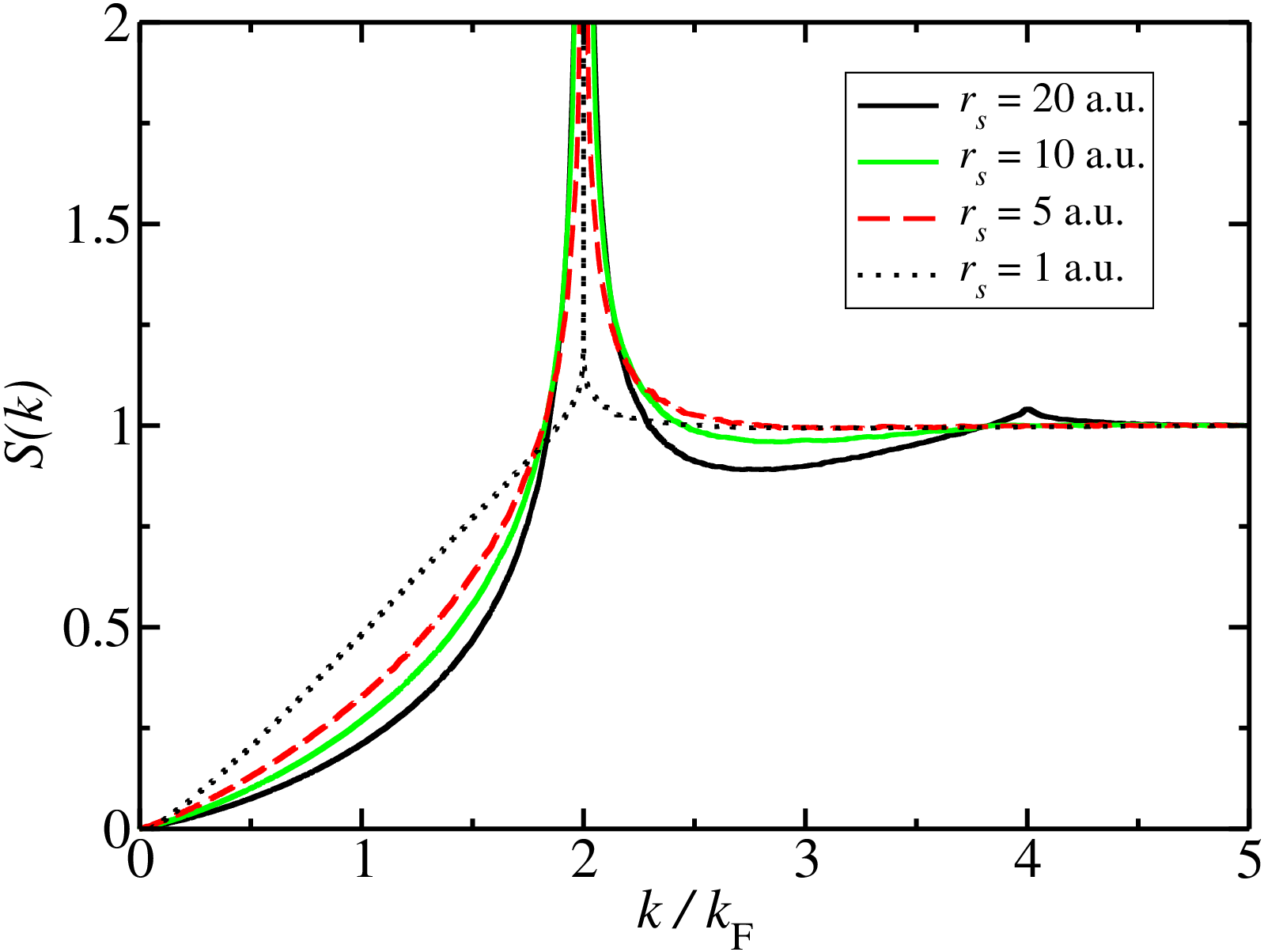}
    \caption{(Color online) SSFs of infinitely thin wires obtained from Eq.\
      (\ref{eq:ssf}) and the extended $N=99$ PCFs.
      \label{fig:ssfs_n99_all_densities.eps}}
  \end{figure}
\end{center}
\begin{center}
  \begin{figure}
    \includegraphics[scale=0.31]{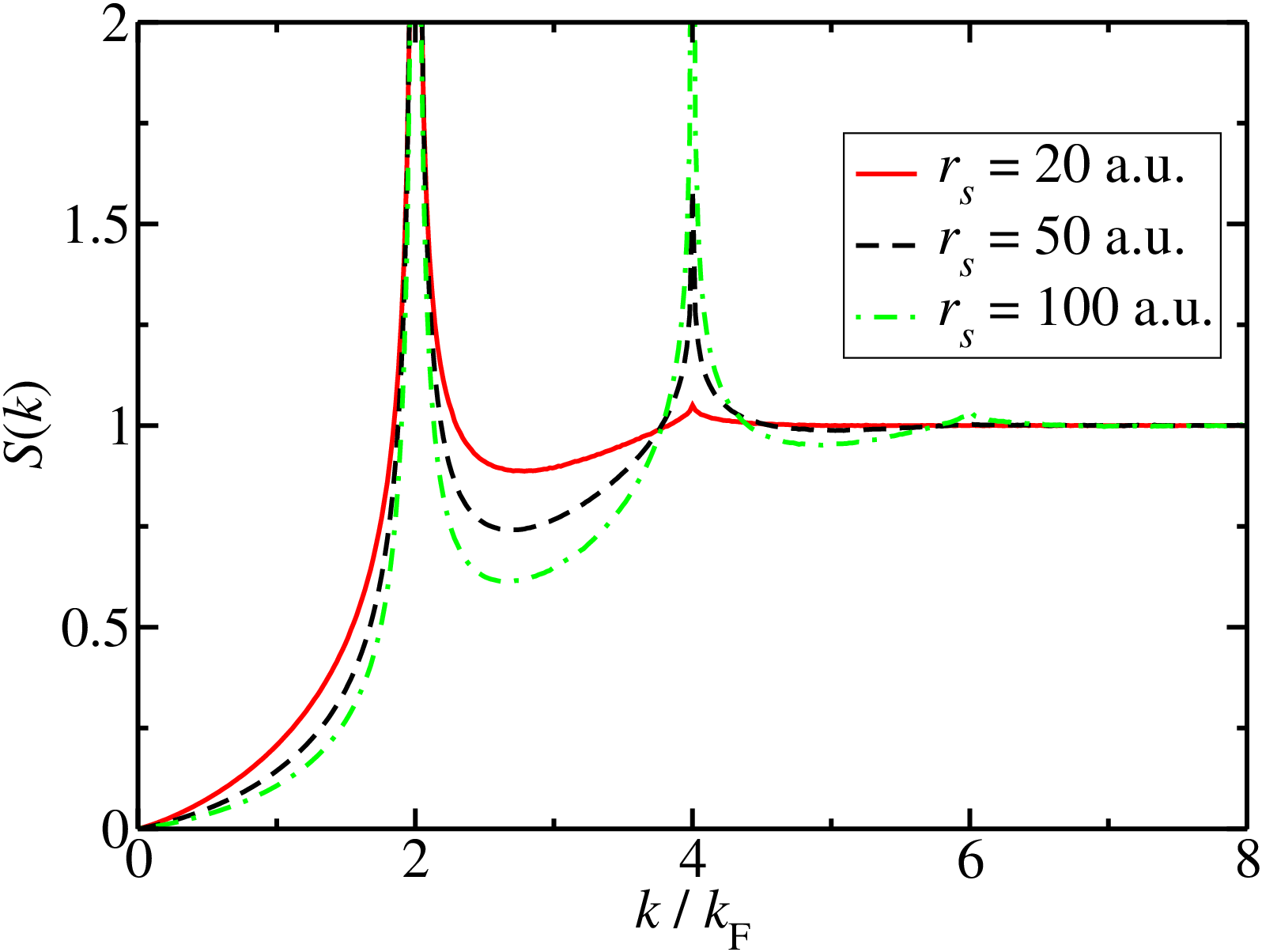}
    \caption{(Color online) SSF of the infinitely-thin wire at very low
      densities. The results shown are for $N=99$.  The finite-cell VMC PCFs
      were used to generate the SSFs in the plot.\label{fig:ultralow_density}}
  \end{figure}
\end{center}
The SSF of the 1D HEG is defined as\cite{Giuliani_2005}
\begin{equation}
  S(k)=1+\frac{N}{L}\int [g(x)-1]e^{-ikx}\;dx\;,
\label{eq:ssf}
\end{equation}
and SSFs that we present here are for the ferromagnetic infinitely-thin
wire. As explained in the introduction, the antiferromagnetic and
ferromagnetic phases are degenerate for the infinitely-thin wire, so we do not
violate the Lieb-Mattis theorem with our choice of system.

The PCF can only be directly measured in QMC for $x<L/2$ due to the
finite extent of the simulation cell. This manifests itself as a
scaling of the SSF peak at $2k_{\mathrm F}$ with system size, as
demonstrated by the data in Fig.\ \ref{fig:rs02_ssf_plot}. The height
of the $2k_{\mathrm F}$ peak in the finite-cell SSFs does not scale as
$N$ (and so $L$) to any single power but appears to be sub-linear,
consistent with the presence of quasi long-range order. At $k$ away
from the peak the SSFs appear to agree very well for different cell
sizes.

We further investigated finite-size effects by performing a fit to the
oscillatory tails of the PCF and using the fitted function to extend the PCF
far beyond $L/2$ before using Eq.\ (\ref{eq:ssf}) to calculate the
SSF\@. After testing a number of functional forms, we found that a
good-quality and simple fit to the oscillatory tails of the PCF takes the
form\cite{Casula_2006,Schulz_1993}
\begin{equation}
  g(x)-1 =A\cos (2k_{\mathrm F} x)\exp (-B\sqrt{\ln x})\;,
\label{eq:pcf_tails}
\end{equation}
where $A$ and $B$ are treated as fitting parameters. The choice of Eq.\
(\ref{eq:pcf_tails}) is motivated by the charge-charge correlation function of
Ref.\ \onlinecite{Schulz_1993}. The parameters we obtained when fitting Eq.\
(\ref{eq:pcf_tails}) to our results are given in Table \ref{table:pcf_fit} in
Appendix \ref{app:pcf_fit}.

We fitted Eq.\ (\ref{eq:pcf_tails}) to the PCF data for ${6r_s<x<L/2-6r_s}$,
although we found that the results were not very sensitive to the region of
data included in the fit. The data close to the origin were not included in
the fit since Eq.\ (\ref{eq:pcf_tails}) is only a good fit for long-range
correlations. The data at the edge of the cell were excluded because that is
the region midway between the electron at the origin and its next periodic
image, and might be expected to be a region where the PCF suffers particularly
badly from finite-size effects.

We then formed the extended PCF by reinstating all of the original PCF data up
to $L/2-6r_s$ and appending a tail for $x>L/2-6r_s$ using Eq.\
(\ref{eq:pcf_tails}) and the fitted parameters. Performing the Fourier
transform of Eq.\ (\ref{eq:ssf}) numerically on the extended PCF results in a
SSF (for $r_s\leq 20$ a.u.)\ with a greatly-enhanced peak at $2k_{\mathrm F}$,
but that agrees very well with the finite-cell SSF everywhere else. Figure
\ref{fig:rs20_pcf_extended.eps} shows the difference between the SSFs obtained
from the finite-cell and the extended PCFs. Under the extension scheme, the
peak at $2k_{\mathrm F}$ appears to be susceptible to noise; in particular,
the density of $k$-points at which the SSF is calculated heavily affects the
apparent height.  The fitting function of Eq.\ (\ref{eq:pcf_tails}) possesses
a peak at $2k_{\mathrm F}$ in Fourier space and smoothly decays away
elsewhere, and is expected to be closer to the $N=\infty$ limit.

The asymptotically-correct charge-charge correlation function of
Schulz\cite{Schulz_1993} includes higher order terms containing
oscillations at wave numbers given by even multiples of $k_{\mathrm
  F}$. For $r_s<15$ a.u.\ we find no discernable features at larger
$k$. However, a small feature at $4k_{\mathrm F}$ starts to develop at
$r_s\approx 15$ a.u., and for $r_s=20$ a.u.\ we observe a clear peak,
visible in Figs.\ \ref{fig:rs20_pcf_extended.eps},
\ref{fig:ssfs_n99_all_densities.eps}, and
\ref{fig:ultralow_density}. We performed short VMC calculations at
extremely low densities, $r_s=50$ a.u.\ and $100$ a.u., where the
electron-electron coupling is very large, to search for more
noticeable features at $k>2k_{\mathrm F}$. We find that peaks in the
SSF do indeed appear at even multiples of $k_{\mathrm F}$ for these
systems, as evidenced in Fig.\ \ref{fig:ultralow_density}. The SSF of
the $r_s=100$ a.u.\ system has clear peaks at $k=2k_{\mathrm F}$,
$4k_{\mathrm F}$, and $6k_{\mathrm F}$. This suggests that one could
add higher order terms to the fit of Eq.\ (\ref{eq:pcf_tails}) for the
low density systems and perform the extension scheme again, although
this seems unlikely to produce any new behavior.

% Repeating the extension procedure by regenerating PCF data using
% different random seeds informs us that the noise in the peak is
% approximately $1\%$ of the peak height.

\subsection{Momentum density\label{sec:md}}
\begin{center}
  \begin{figure}
    \includegraphics[scale=0.31]{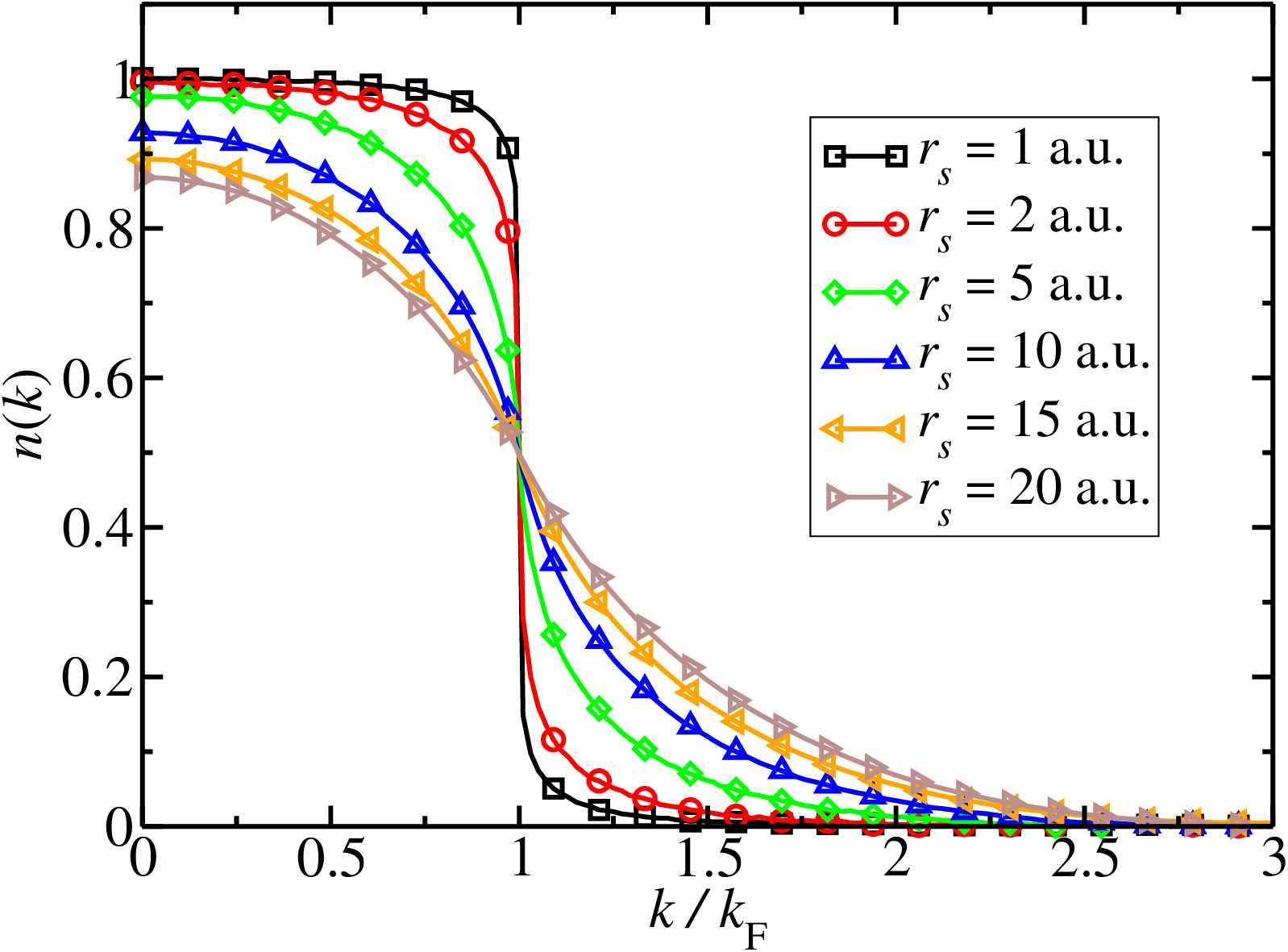}
    \caption{(Color online) MD of the infinitely-thin wire
      at several densities. The data shown are for $N=99$ and are extrapolated
      estimates ${[2n_{\rm DMC}(k)-n_{\rm VMC}(k)]}$. The statistical error
      bars are much smaller than the symbols and some symbols have been
      omitted for clarity.\label{fig:md_plot}}
  \end{figure}
\end{center}
\begin{center}
  \begin{figure}
    \includegraphics[scale=0.31]{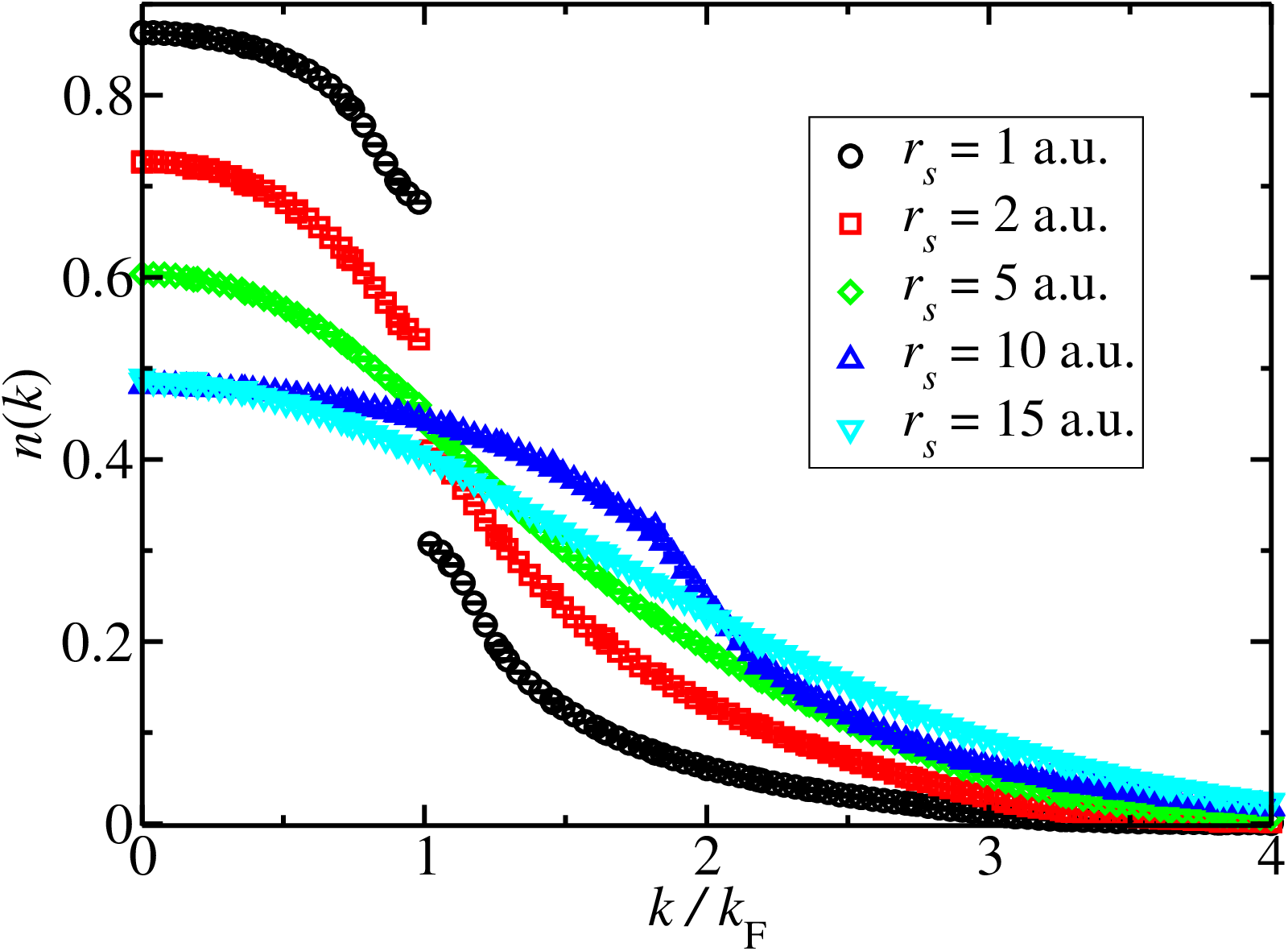}
    \caption{(Color online) VMC MD of the harmonic wire with $b=0.1$
      a.u.\ and $\zeta=0$ at several densities. The data shown for
      each density are for $N=22$ and $102$ (joined to form one data
      set). The statistical error bars are smaller than the
      symbols.\label{fig:harmwire_md_plot}}
  \end{figure}
\end{center}
\begin{center}
  \begin{figure}
    \includegraphics[scale=0.31]{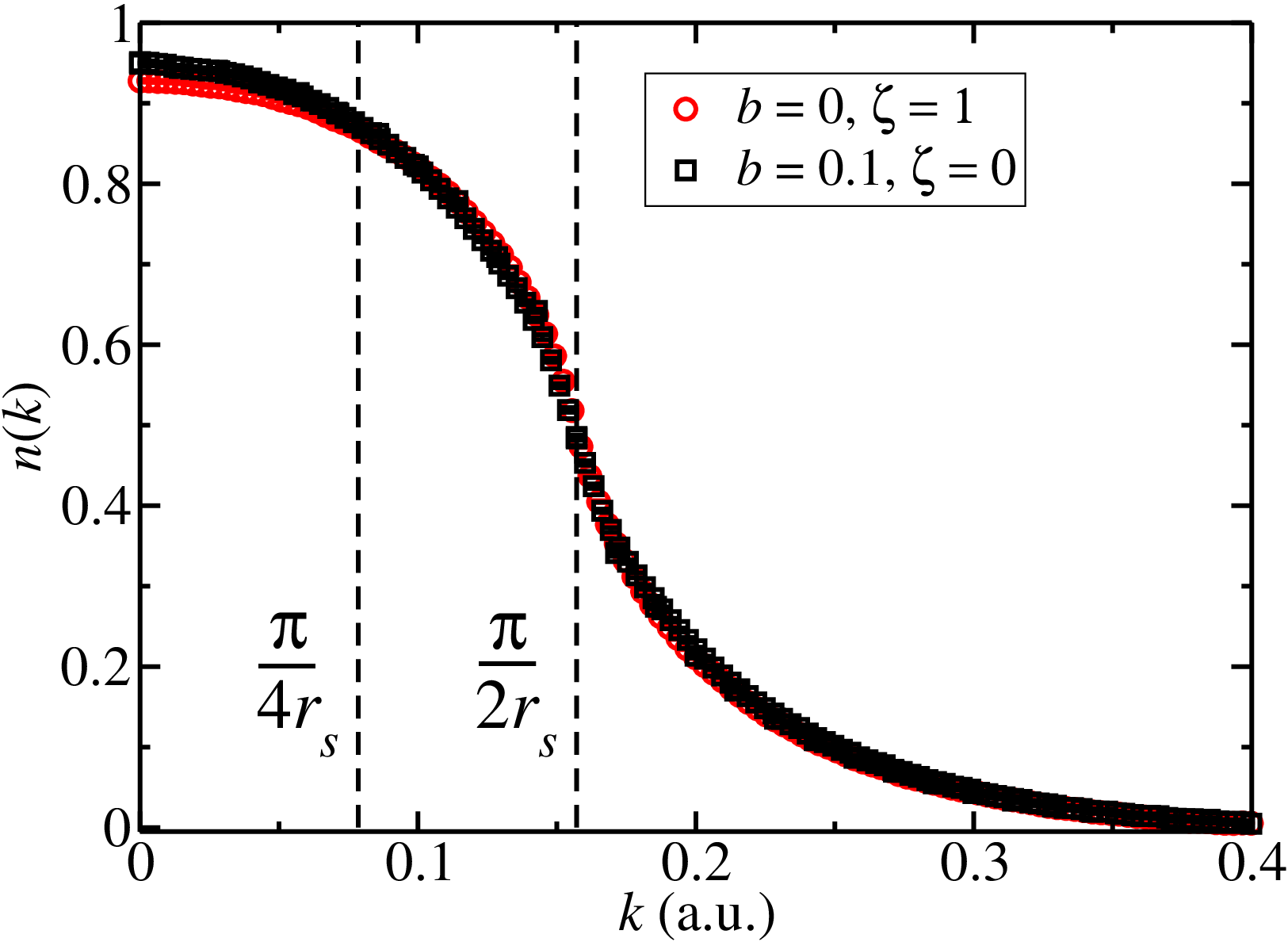}
    \caption{(Color online) Comparison of the $r_s=10$ a.u.\ MD for the
      infinitely-thin wire ($b=0$, $\zeta=1$) with that obtained for the
      harmonic wire with $b=0.1$ a.u.\ and $\zeta=0$. The statistical error
      bars are similar in size to the symbols. The dashed lines show the
      values of $k$ at which Eq.\ (\ref{eq:exponentfit}) was fitted to the
      data for the calculations of the exponent
      $\alpha$.\label{fig:rs10_md_plot}}
  \end{figure}
\end{center}
The MD is accumulated as
\begin{equation}
  n(k)=\left < \frac{1}{2\pi}\int \frac{\Psi_{\mathrm T}(r)}{\Psi_{\mathrm
    T}(x_1)}\exp[ik (x_1-r)]\;dr  \right >\;,
  \label{eq:momden}
\end{equation}
where $\Psi_{\mathrm T}(r)$ is the trial wave function evaluated at
$(r,x_2,\ldots,x_n)$ and angular brackets denote an average over
configurations. The MD is the integral of the spectral function from
minus infinity up to the chemical potential.\cite{Giuliani_2005} The
MD exhibits a drop at $k_{\mathrm F}$ because the peak in the spectral
function reaches and passes through the chemical potential. If the
peak in the spectral function is a $\delta$-function at $k_{\mathrm
  F}$ (\textit{i.e.,} the spectral function possesses a quasiparticle
peak) then the MD is discontinuous at the Fermi edge. However, in 1D
we expect the excitations to be collective rather than
single-particle-like. The 1D systems should thus have MDs that are
continuous at $k_{\mathrm F}$, although TL liquid theory predicts that
the gradient will be singular.\cite{Schulz_1993}

For the systems with $\zeta=1$, we have used $k_{\mathrm
  F}=\pi/(2r_s)$, whereas for the systems with $\zeta=0$, we have used
$k_{\mathrm F}=\pi/(4r_s)$. Figure \ref{fig:md_plot} shows the MDs
that we obtain by evaluating the extrapolated estimator
$2n_{\mathrm{DMC}}(k)-n_{\mathrm{VMC}}(k)$ for the infinitely-thin
wire. The VMC and DMC results differed by no more than $\sim 2$ error
bars, so that evaluating the extrapolated estimator changed the
results very little. Figure \ref{fig:harmwire_md_plot} shows the MD
for the harmonic wire with $b=0.1$ a.u.\ and $\zeta=0$.

A particularly interesting feature of the paramagnetic harmonic wire
MD is that as $r_s$ is increased and $b$ is decreased the function
shifts much of its weight to larger $k$, and $n(0)$ reduces to values
around $0.5$. This is a direct manifestation of the harmonic wire
becoming more like the ferromagnetic infinitely-thin system. One can
in some cases see a feature resembling the gradient discontinuity
appearing at $\pi/(2r_s)$, \textit{i.e.,} at twice the paramagnetic
Fermi wave vector. In particular, for $r_s=10$ a.u.\ and $b=0.1$ a.u.\
the MD possesses a feature at $\pi/(2r_s)$. Upon closer inspection we
find that the MD for the unpolarized system with $b=0.1$ a.u.\ agrees
very well with that of the infinitely-thin wire ($b=0$ and
$\zeta=1$). Figure \ref{fig:rs10_md_plot} illustrates this
comparison. It thus appears possible to in some sense tune the
effective Fermi wave vector by adjusting the strength of the
confinement (and the density). A dense paramagnetic system with very
weak confinement shows significant occupation of momentum states up to
approximately $\pi/(4r_s)$. Increasing the effective interaction
strength increases this value of $k$ until it eventually saturates at
the ferromagnetic $k_F$. This reflects the fact that in the limit
$r_s\rightarrow \infty$ the pseudonodes at antiparallel-spin
coalescence points become true nodes.

\subsection{Tomonaga-Luttinger liquid parameters\label{sec:ll_params}}
\begin{center}
  \begin{figure}
    \includegraphics[scale=0.31]{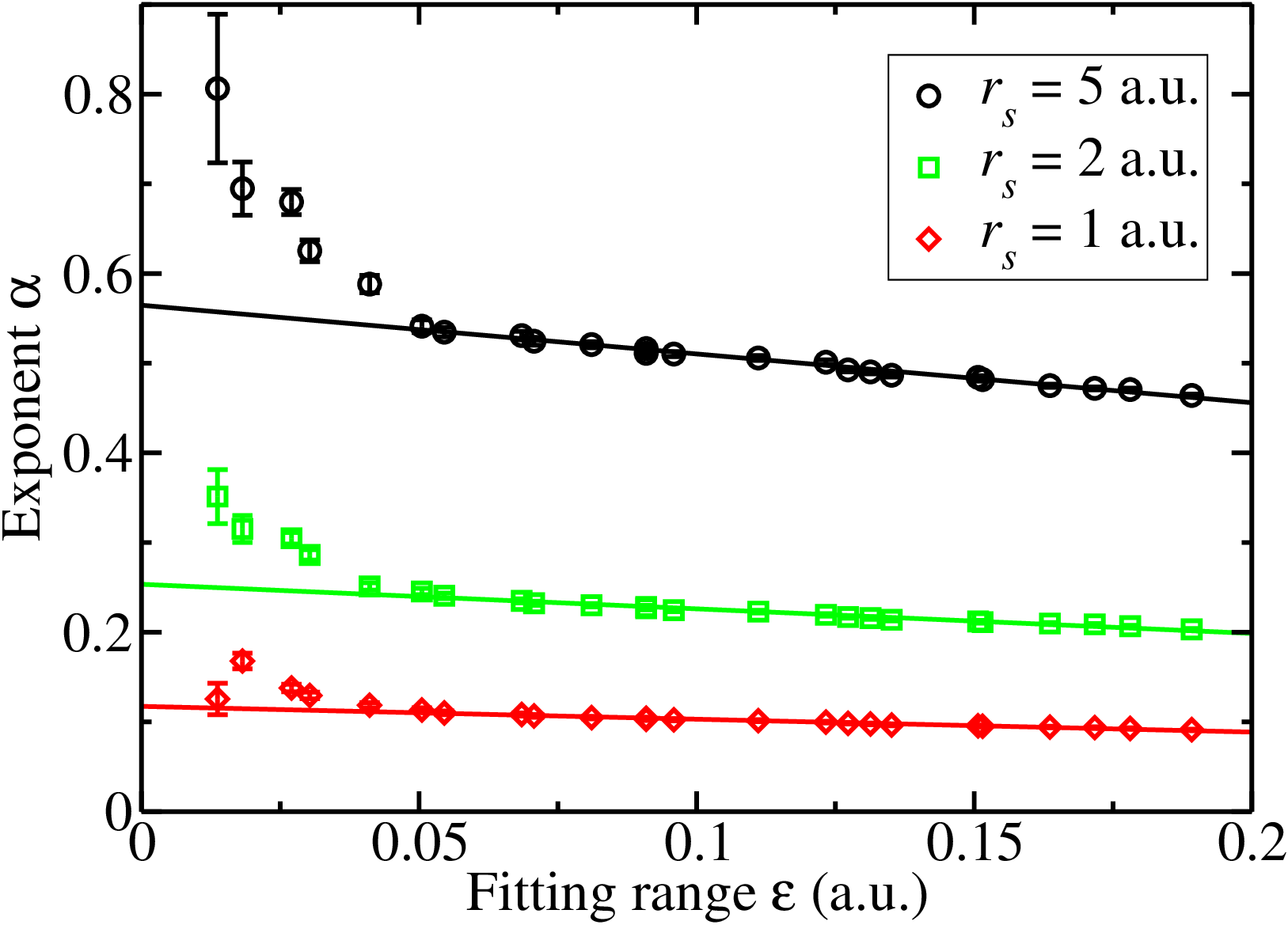}
    \caption{(Color online) Exponent $\alpha$ [found from fitting Eq.\
      (\ref{eq:exponentfit}) to the MD] against the range of
      data included in the fit. The range of data is described by
      $|k-k_{\mathrm F}|<\varepsilon k_{\mathrm F}$. The symbols are the
      fitted exponent values and the solid lines are linear fits to the
      exponents in the region $\varepsilon>0.05$. The data shown are for the
      infinitely-thin wire. \label{fig:exponent_extrap}}
  \end{figure}
\end{center}
\begin{center}
  \begin{figure}
    \includegraphics[scale=0.31]{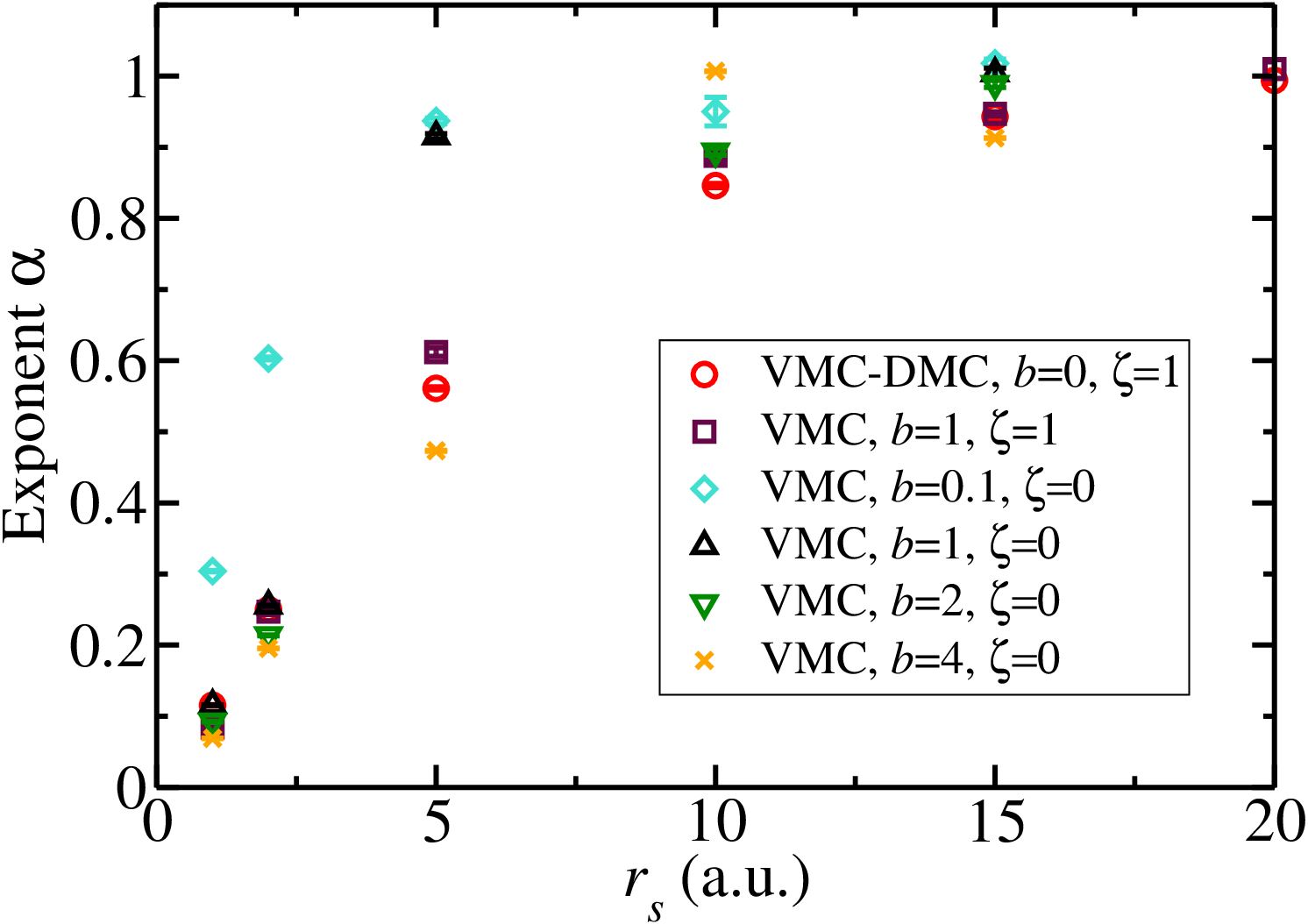}
    \caption{(Color online) Exponent $\alpha$ found from fitting
      Eq.\ (\ref{eq:exponentfit}) to the MD around $k=\pi/(2r_s)$
      for the $\zeta=1$ systems and $k=\pi/(4r_s)$ for the $\zeta=0$
      systems. \label{fig:exponents}}
  \end{figure}
\end{center}
\begin{center}
  \begin{figure}
    \includegraphics[scale=0.31]{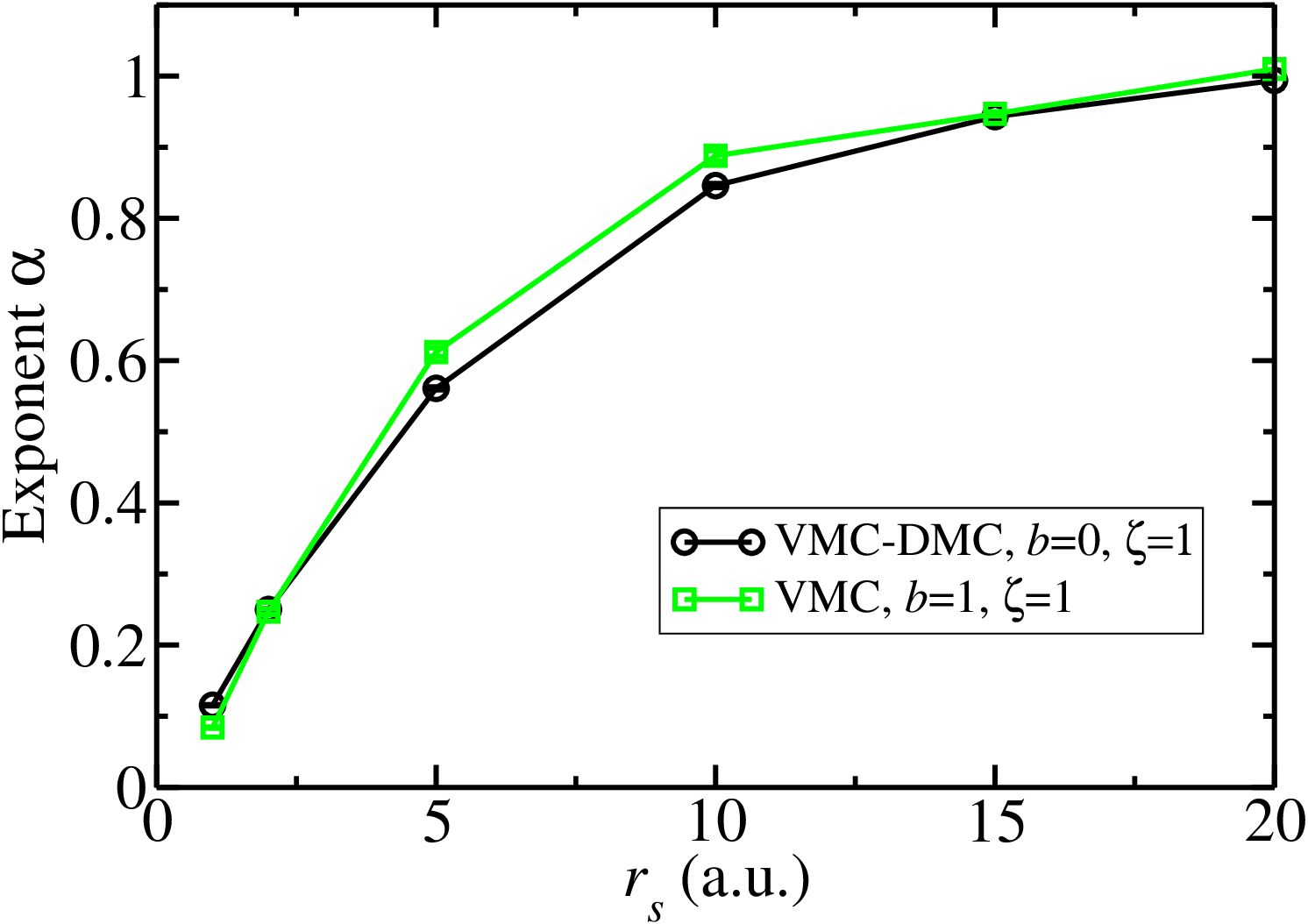}
    \caption{(Color online) Exponent $\alpha$ found from fitting
      Eq.\ (\ref{eq:exponentfit}) to the MDs of the ferromagnetic
      systems. \label{fig:exponents_zeta1}}
  \end{figure}
\end{center}
\begin{center}
  \begin{figure}
    \includegraphics[scale=0.31]{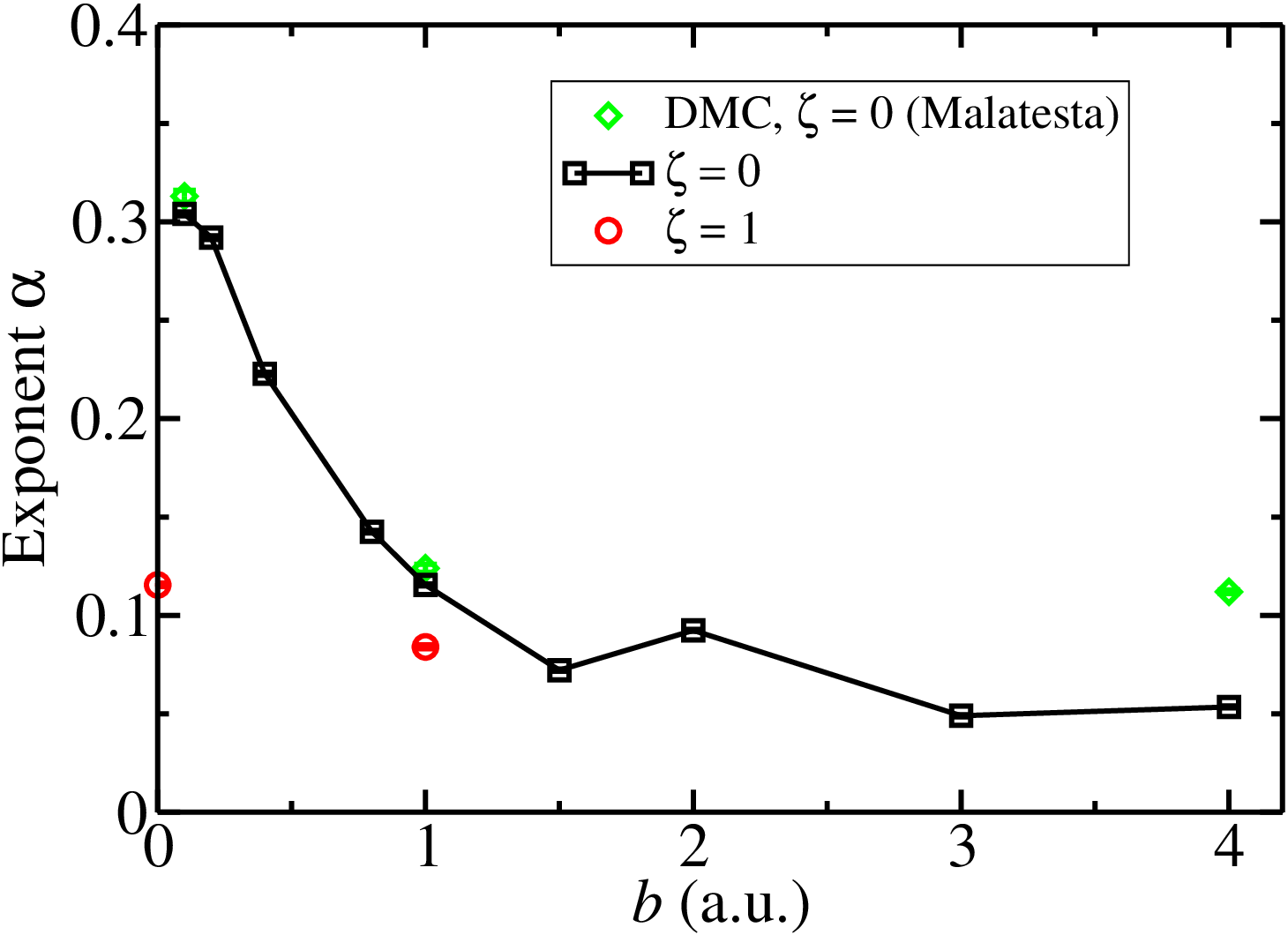}
    \caption{(Color online) Exponent $\alpha$ found from fitting
      Eq.\ (\ref{eq:exponentfit}) to the MD of systems with
      $r_s=1$ a.u. \label{fig:rs1_exponents}}
  \end{figure}
\end{center}
Close to the Fermi wave vector, TL liquid theory suggests that the
MD takes the form\cite{Luttinger_1963,Mattis_1965}
\begin{equation}
  \label{eq:exponentfit}
  n(k)=n(k_{\mathrm F})+A[{\rm sign}(k-k_{\mathrm F})]|k-k_{\mathrm
  F}|^{\alpha}\;,
\end{equation}
which we have fitted to our results treating $n(k_{\mathrm F})$, $A$, and
$\alpha$ as fitting parameters.  Note that within TL liquid theory the
exponent $\alpha$ is related to the TL liquid parameter\cite{Schulz_1990}
$K_\rho$ by
\begin{equation}
  \label{eq:luttinger_parameter}
  \alpha=\frac{1}{4}\left ( K_\rho + \frac{1}{K_\rho}-2\right )\;.
\end{equation}

If the range of data included in the fit is described by
$|k-k_{\mathrm F}|<\varepsilon k_{\mathrm F}$, the choice of
$\varepsilon$ can present some difficulties. Ideally, one would choose
$\varepsilon \rightarrow 0$ since Eq.\ (\ref{eq:exponentfit}) is
potentially valid for $k\rightarrow k_{\mathrm F}$, and indeed using
the entire range of MD results yields rather poor fits. However, the
estimate of $\alpha$ becomes noisy when $\varepsilon$ is small, and at
the extreme where just two data points are included, one can of course
obtain any value for $\alpha$. This leads us to include fits
constructed using a larger range of $k$ values. In practice, we chose
to perform a linear extrapolation to $\varepsilon=0$ excluding fits
where $\varepsilon<0.05$ for the $\zeta=1$ systems, as shown in Fig.\
\ref{fig:exponent_extrap}. For the $\zeta=0$ systems, the
extrapolation used $\alpha$ from fits for which $\varepsilon\gtrsim
0.25$. The trend that we observe in the exponent with respect to
$\varepsilon$ is similar to that found in Ref.\
\onlinecite{Wang_2001}.

% figure 1 of that reference looks very similar to fig:exponent_extrap

Figure \ref{fig:exponents} shows the exponents $\alpha$ that we obtain
for several densities, polarizations, and confinements. All of the
systems show the same general trend; $\alpha$ tends to $0$ in the
high-density limit and to $1$ in the low-density limit. As mentioned
earlier, it is important to note that for the $\zeta=1$ systems we
fitted Eq.\ (\ref{eq:exponentfit}) to the MD at $\pi/(2r_s)$, whereas
for $\zeta=0$ we used $\pi/(4r_s)$. The change in shape of the MD upon
varying the interaction strength that we noted in Sec.\ \ref{sec:md},
and the apparent shift in $k_{\mathrm F}$, suggests that one could
also extract a relevant exponent from fitting to other values of
$k$. For example, we showed in Fig.\ \ref{fig:rs10_md_plot} the
similarity between the $r_s=10$ a.u., $b=0.1$ a.u., $\zeta=0$ MD and
that of the $r_s=10$ a.u., $b=0$, $\zeta=1$ system. Despite the
similarity of the MDs for the two systems, the fits used to extract
$\alpha$ from each were performed at different values of $k$ --- a
factor of two apart in fact. The result is that the exponent for the
paramagnetic wire is larger, since the Fermi edge for that system has
apparently shifted to $k$ above the fitting region.

With this in mind, Fig.\ \ref{fig:exponents_zeta1} shows the $\zeta=1$ results
alone, since for the ferromagnetic systems one can clearly and reliably state
that $k_{\mathrm F}=\pi/(2r_s)$ for the whole range of densities. The exponent
$\alpha$ for the infinitely-thin wire is reasonably well-approximated by the
function
\begin{equation}
  \label{eq:param_fit}
  \alpha=\tanh (r_s / 8)\;,
\end{equation}
which gives a maximum deviation of $0.011(3)$ from the $b=0$ QMC results,
which occurs at $r_s=15$ a.u. The exponents for the harmonic wire with $b=1$
a.u.\ and $\zeta=1$ show a maximum deviation from Eq.\ (\ref{eq:param_fit}) of
$0.057(6)$, which we find at $r_s=5$ a.u.

The exponent $\alpha$ has been reported in previous theoretical and
experimental studies. Reference \onlinecite{Malatesta_thesis} gives the
exponents for $b=0.1$, $1$, and $4$ a.u.\ (with $r_s=1$ a.u.\ and $\zeta=0$)
from VMC calculations. In Fig.\ \ref{fig:rs1_exponents} we have shown how the
results given there compare with ours. It appears that the principal
difference between the two studies is the procedure for deciding upon a
fitting region; Ref.\ \onlinecite{Malatesta_thesis} does not give details of
any extrapolation to $\varepsilon=0$ and presumably the whole range of $n(k)$
was included in the fit. Figure \ref{fig:rs1_exponents} also includes the
exponent we find for the infinitely-thin wire (from VMC and DMC estimates of
the MD) at $b=0$.

The exponent $\alpha$ has also been reported from experiments, mostly through
measurements accessing the single-particle density of states near the Fermi
edge. The exponent for carbon nanotubes ranges between $0.2$ and $0.4$,
although it is difficult to map the behavior of electrons in these systems
onto our model since the electronic properties depend on the folding
geometry.\cite{Dresselhaus_2001,Egger_1998,Bockrath_1999,Ishii_2003,Shiraishi_2003}
For the Bechgaard salts, which have a 1D carrier density of $r_s\approx 6.9$
a.u., exponents between $0.5625$ and $0.8$ have been
reported.\cite{Popovic_2001,Vescoli_1998,Dressel_2005,Georges_2000}

\section{Conclusions\label{sec:conclusions}}
We have presented calculations of the ground state energy, PCF, SSF, and MD of
the infinitely-thin 1D HEG model using VMC and DMC\@. We observe the
development of peaks at increasingly-large even-integer-multiples of
$k_{\mathrm F}$ in the SSF as the density is lowered, consistent with the
predictions of Schulz.\cite{Schulz_1993}

For the harmonic wire model, we have reported ground-state MDs and TL
parameters for a range of densities and confinements. We used VMC to
produce these results; comparison of our PCFs, SSFs, and ground state
energies with LRDMC results\cite{Casula_2006} where available
indicates that our results are extremely accurate.

The MDs of the $\zeta=0$ systems tend towards the MDs
of the infinitely-thin wire and ferromagnetic harmonic wire as $b$ is
decreased and as $r_s$ is increased, both of which have the effect of
increasing the electron-electron coupling. One interpretation for this is that
correlation is dominating over kinetic confinement, so that antiparallel spin
pairs are avoiding one another almost as much as parallel spin pairs.

The TL parameters calculated for the $b=1$ a.u., $\zeta=1$ system show
reasonable agreement with the infinitely-thin wire results; the
maximum deviation that we observe between the parameters for the two
systems is $0.051(6)$, which occurs at $r_s=5$ a.u. The exponent
$\alpha$, which describes the behavior of the MD at $k_{\mathrm F}$,
takes values between $0$ and $1$. The exponent for the $\zeta=0$
systems shows the same general trend, although the value of $\alpha$
is typically higher than for the $\zeta=1$ systems. This is largely a
consequence of the shift of the weight in the MD (including the
singularity in the gradient) to larger $k$ as the coupling is
increased.

%; we held the fitting region at $k=\pi/(4r_s)$ when $\zeta=0$.

\begin{acknowledgments}
  We acknowledge financial support from the U.K.\ Engineering and Physical
  Sciences Research Council (EPSRC)\@. Computing resources were provided by
  the Cambridge HPCS\@.  We would like to thank Richard Needs for many helpful
  discussions. We would also like to thank M.\ Casula for sending us his LRDMC
  data for the harmonic wire and G.\ Senatore for sending us VMC results for
  the harmonic wire.
\end{acknowledgments}

\appendix

\section{Derivation of the quasi-1D interaction\label{app:ewald}}
One may derive Eq.\ (\ref{eq:Ewald_like_sum}) from first principles.
Suppose the energy scales are such that we may write the wave function
as a product $\theta(\vect{r}_\perp)\psi(x)$, where $x$ is the
projection of the electron position onto the axis of the wire and
$\vect{r}_\perp$ is the transverse position.

If the electrons are sufficiently confined in the transverse plane, one may
obtain the 1D interaction $v(x)$ by integrating over the transverse part of
the wave function,
\begin{equation}
  v(x)=\int\frac{|\theta(\vect{r}_{\perp})|^2\;
  |\theta(\vect{r}'_{\perp})|^2}{\big
  [x^2+|\vect{r}_{\perp}-\vect{r}'_{\perp}|^2\big ]^{1/2}}\;
  d\vect{r}_{\perp} d\vect{r}'_{\perp}\;.
\label{eq:e-e_int_simple}
\end{equation}
For the harmonic wire, the confining potential is $r^2_{\perp}/8b^4$, where
$b$ is a parameter. If $r_s \gg \pi b /4$, one may make the assumption that
the electrons occupy only the lowest sub-band, which is given by
\begin{equation}
  \label{eq:transverse_wfn}
  \theta(r_\perp)=\frac{1}{\sqrt{2\pi b^2}}\exp\left (
  -\frac{r_\perp^2}{4b^2}\right )\;.
\end{equation}
Substituting Eq.\ (\ref{eq:transverse_wfn}) into Eq.\
(\ref{eq:e-e_int_simple}) yields\cite{Friesen_1980}
\begin{equation}
v(x)=\frac{\sqrt{\pi}}{2b}\exp \left ( \frac{x^2}{4b^2} \right ) {\rm erfc}
\left ( \frac{|x|}{2b} \right ) \;,
\label{eq:e-e_int_full}
\end{equation}
which is finite at $x=0$ but retains a long-range $1/|x|$ tail. The Fourier
transform of Eq.\ (\ref{eq:e-e_int_full}) is
\begin{equation}
\tilde{v}(k)=E_1(b^2k^2)\exp (b^2k^2)\;,
\label{eq:FT_e-e_int}
\end{equation}
where $E_1$ is the exponential integral function.

Having found the real and reciprocal space representations of the 1D
interaction in a harmonic wire, one must perform an Ewald-like sum to enable
calculations with periodic systems. We follow a route similar to that of Ref.\
\onlinecite{Malatesta_thesis}.

The interaction of an electron at the origin with another at position $x$, all
of that electron's periodic images, and its background is given by
\begin{equation}
\phi(x)=\sum_{m=-\infty}^\infty \Big\lbrace
v(x-mL)-\frac{1}{L}\int^{L/2}_{-L/2}dy\;v(x-mL-y)\Big\rbrace\;,
\label{eq:Ewald_simple}
\end{equation}
where $L$ is the length of the simulation cell. The objective is to write Eq.\
(\ref{eq:Ewald_simple}) in terms of quickly converging discrete sums. The
first step is to write Eq.\ (\ref{eq:Ewald_simple}) in the more useful form
\begin{eqnarray}
\phi(x)=\gamma_0(x) -\frac{1}{L}\int_{-\infty}^{\infty}dy\;v(x-y)\;,
\label{eq:Ewald_rewrite}
\end{eqnarray}
where
\begin{equation}
\gamma_0(x)=\sum_{m=-\infty}^\infty v(x-mL)\;.
\label{eq:gamma0}
\end{equation}
Equation (\ref{eq:gamma0}) is already in a form that is quick and easy to
evaluate, so we turn our attention to reformulating the integral in the second
term of Eq.\ (\ref{eq:Ewald_rewrite}). We first perform the trick of both
adding and subtracting a Gaussian term $p(y)$, giving
\begin{equation}
-\frac{1}{L}\int_{-\infty}^{\infty}dy\;v(x-y)=\gamma_1(x)+\gamma_2(x)\;,
\label{eq:Ewald_gammas_rewrite}
\end{equation}
where
\begin{eqnarray}
\gamma_1(x)&=&-\int^{\infty}_{-\infty}dy\;v(x-y) p(y) \;,
\label{eq:gamma1} \\
\gamma_2(x)&=&\int^{\infty}_{-\infty}dy\;v(x-y) \left [p(y)
  -\frac{1}{L}\right ]\;,
\label{eq:gamma2}
\end{eqnarray}
and the term that we have added and subtracted is
\begin{equation}
p(y)=\sum_{m=-\infty}^{\infty} \frac{1}{2b\sqrt{\pi}}\exp \left (
-\frac{1}{4b^2} (y-mL)^2\right )\;.
\label{eq:trick_gaussian}
\end{equation}

It is clear that $\phi(x)$ may now be written simply as
\begin{equation}
\phi(x)=\gamma_0(x)+\gamma_1(x)+\gamma_2(x)\;.
\label{eq:phi_sum_gammas}
\end{equation}
We first inspect $\gamma_1(x)$, finding that it may be integrated directly to
give
\begin{equation}
\gamma_1(x)=\sum_{m=-\infty}^{\infty}\left \lbrace -\frac{1}{|x-mL|}\;{\rm
  erf}\left (  \frac{|x-mL|}{2b}\right ) \right \rbrace\;,
\label{eq:gamma1_final}
\end{equation}
which is a form suitable for numerical evaluation.

One may take the first step towards simplifying $\gamma_2(x)$ by performing a
Poisson summation on $p(y)$,
\begin{equation}
p(y)=\frac{1}{L}\left [ 1+2\sum^\infty_{n=1}e^{-(bGn)^2}\cos (Gn y) \right ]\;,
\label{eq:Fourier_Gaussian}
\end{equation}
where $G=2\pi/L$. Putting Eq.\ (\ref{eq:Fourier_Gaussian}) into Eq.\
(\ref{eq:gamma2}) gives
\begin{equation}
\gamma_2(x)= \frac{2}{L}\sum^\infty_{n=1}e^{-(bGn)^2}
\int^{\infty}_{-\infty}dy\; v(x-y)\cos (Gn y)\;,
\label{eq:gamma2new}
\end{equation}
which may straightforwardly be rewritten in its final form,
\begin{equation}
\gamma_2(x)=\frac{2\sqrt{2\pi}}{L}\sum_{n=1}^\infty \tilde{v}(Gn)\cos(Gnx)
\;e^{-(bGn)^2}\;,
\label{eq:gamma2_final}
\end{equation}
where we have used the result
\begin{equation}
  \int^{\infty}_{-\infty}dy\;v(x-y)\cos (Gn y)=
  \sqrt{2\pi}\;\tilde{v}(Gn)\cos(Gnx)\;.
\label{eq:gamma2new2}
\end{equation}

Finally, putting the expressions for the $\gamma$ functions, Eqs.\
(\ref{eq:gamma0}), (\ref{eq:gamma1_final}), and (\ref{eq:gamma2_final}), into
Eq.\ (\ref{eq:phi_sum_gammas}) and remembering that $\tilde{v}(k)$ is given by
Eq.\ (\ref{eq:FT_e-e_int}), we obtain the more computationally convenient form
\begin{eqnarray}
  \phi(x)&=& \sum_{m=-\infty}^{\infty}\Bigg \lbrace \frac{\sqrt{\pi}}{2b}
  e^{(x-mL)^2/(4b^2)} {\rm erfc} \left (
  \frac{|x-mL|}{2b}\right )\nonumber \\ &-&\frac{1}{|x-mL|} \; {\rm  erf}
  \left (  \frac{|x-mL|}{2b}\right ) \Bigg \rbrace\nonumber \\
  &+&\frac{2}{L}\sum_{n=1}^\infty  E_1 \left [(bGn)^2 \right ]\cos(Gnx)\;.
\label{eq:derivation_end}
\end{eqnarray}
It should be noted that in Ref.\ \onlinecite{Casula_2006}, Rydberg rather than
Hartree units were used so that the potentials given there differ from Eq.\
(\ref{eq:derivation_end}) by a factor of 2.
\section{Pair correlation function fitting parameters\label{app:pcf_fit}}
Table \ref{table:pcf_fit} shows the parameters that we obtained when fitting
Eq.\ (\ref{eq:pcf_tails}) to the extrapolated estimates of the PCF for the
infinitely-thin wire. We performed the fit for $r_s=1$, $2$, $5$, $10$, $15$,
and $20$ a.u.\ with systems containing $N=37$, $55$, $73$, and $99$
particles. The PCF data in the range ${6r_s<x<L/2-6r_s}$ were included in the
fit.
\begin{table}
  \begin{center}
   \vspace{0.5cm}
   \begin{tabular}{l@{~~~} lr@{.}lr@{.}l}
     \hline\hline

     $r_s$ (a.u.) & \;$N$\;\;\;\; & \multicolumn{2}{c}{$A$}& \multicolumn{2}{c}{$B$ (a.u.)} \\ \hline

1 & 37 & 1&908\hspace{14pt} & 3&291 \\

1 & 55 & 3&090 & 3&683 \\

1 & 73 & 1&940 & 3&446 \\

1 & 99 & 2&113 & 3&671 \\

2 & 37 & 4&851 & 2&952 \\

2 & 55 & 4&573 & 2&979 \\

2 & 73 & 6&047 & 3&069 \\

2 & 99 & 8&545 & 3&273 \\

5 & 37 & 8&029 & 2&237 \\

5 & 55 & 8&310 & 2&258 \\

5 & 73 & 10&061 & 2&359 \\

5 & 99 & 9&262 & 2&320 \\

10 & 37 & 8&465 & 1&735 \\

10 & 55 & 9&349 & 1&788 \\

10 & 73 & 9&066 & 1&780 \\

10 & 99 & 10&206 & 1&839 \\

15 & 37 & 8&520 & 1&502 \\

15 & 55 & 8&363 & 1&501 \\

15 & 73 & 8&918 & 1&534 \\

15 & 99 & 9&788 & 1&580 \\

20 & 37 & 7&754 & 1&320 \\

20 & 55 & 8&361 & 1&359 \\

20 & 73 & 8&625 & 1&377 \\

20 & 99 & 8&895 & 1&396 \\

     \hline \hline
   \end{tabular}
   \caption{Table showing the fitting parameters $A$ and $B$ from Eq.\
     (\ref{eq:pcf_tails}) obtained from fitting to the extrapolated estimates
     of the PCF for the infinitely-thin wire. The fits were to PCF data in the
     range ${6r_s<x<L/2-6r_s}$.
     \label{table:pcf_fit}}
 \end{center}
 \end{table}


\begin{thebibliography}{99}

\bibitem{Giuliani_2005} G.\ F.\ Giuliani and G.\ Vignale, \textit{Quantum
    Theory of the Electron Liquid} (Cambridge University Press, Cambridge,
    2005).

\bibitem{Tomonaga_1950} S.\ Tomonaga, Prog.\ Theor.\ Phys.\ \textbf{5}, 544
  (1950).

\bibitem{Luttinger_1963} J.\ M.\ Luttinger, J.\ Math.\ Phys.\ \textbf{4}, 1154
  (1963).

\bibitem{Haldane_1981} F.\ D.\ M.\ Haldane, J.\ Phys.\ C:\ Solid St.\ Phys.\
  \textbf{14}, 2585 (1981).

\bibitem{Mitin_2009} V.\ Mitin, A.\ Sergeev, M.\ Bell, J.\ Bird, and A.\
  Verevkin, J.\ Phys.:\ Conf.\ Ser.\ \textbf{193}, 012116 (2009).

  % LL behavior in C nanotubes
\bibitem{Bockrath_1999} M.\ Bockrath, D.\ H.\ Cobden, J.\ Lu, A.\ G.\ Rinzler,
  R.\ E.\ Smalley, L.\ Balents, and P.\ L.\ McEuen, Nature \textbf{397}, 598
  (1999).

\bibitem{Voit_1994} J.\ Voit, Rep.\ Prog.\ Phys.\ \textbf{57}, 977 (1994).

\bibitem{Schulz_1998} H.\ J.\ Schulz, G.\ Cuniberti, and P.\ Pieri,
  arXiv:cond-mat/9807366v2 [cond-mat.str-el] (1998).

\bibitem{Meden_2002} V.\ Meden, W.\ Metzner, U.\ Schollw\"ock, and
  K. Sch\"onhammer J.\ Low Temp.\ Phys.\ \textbf{126}, 1147 (2002).

  % Organic conductors showing Luttinger liquid behavior
\bibitem{Ito_2005} T.\ Ito, A.\ Chainani, T.\ Haruna, K.\ Kanai, T.\ Yokoya,
  S.\ Shin, and R.\ Kato, Phys.\ Rev.\ Lett.\ \textbf{95}, 246402 (2005).

  % LL params of some organic conductors
\bibitem{Dressel_2005} M.\ Dressel, K.\ Petukhov, B.\ Salameh, P.\ Zornoza,
  and T.\ Giamarchi, Phys.\ Rev.\ B \textbf{71}, 075104 (2005).

  % Observation of spin-charge separation in 1D organic conductors
\bibitem{Lorenz_2002} T.\ Lorenz, M.\ Hofmann, M.\ Gr\"{u}ninger, A.\
  Freimuth, G.\ S.\ Uhrig, M.\ Dumm, and M.\ Dressel, Nature \textbf{418}, 614
  (2002).

  % LL behavior in Bechgaard salts
\bibitem{Schwartz_1998} A.\ Schwartz, M.\ Dressel, G.\ Gr\"{u}ner, V.\
  Vescoli, L.\ Degiorgi, and T.\ Giamarchi, Phys.\ Rev.\ B \textbf{58}, 1261
  (1998).

  % Evidence for a Tomonaga-Luttinger liquid in the Bechgaard salts
\bibitem{Vescoli_2000} V.\ Vescoli, F.\ Zwick, W.\ Henderson, L.\ Degiorgi,
  M.\ Grioni, G.\ Gruner, and L.\ K.\ Montgomery, Eur.\ Phys.\ J.\ B
  \textbf{13}, 503 (2000).

  % Spin and charge dynamics in Bechgaard salts
\bibitem{Dressel_2001} M.\ Dressel, S.\ Kirchner, P.\ Hesse, G.\ Untereiner,
  M.\ Dumm, J.\ Hemberger, A.\ Loidl, and M.\ Montgomery, Synth.\ Met.\
  \textbf{120}, 719 (2001).

  % Physics of transition metal oxides
\bibitem{Maekawa_2004} S.\ Maekawa, T.\ Tohyama, S.\ E.\ Barnes, S.\ Ishihara,
  W.\ Koshibae, and G.\ Khaliullin, \textit{Physics of Transition Metal
  Oxides:\ v.\ 144}, (Springer, 2004).

  % One dimensional transition metal oxides (study elec.\ struc.\ using x-rays)
\bibitem{Hu_2002} Z.\ Hu, M.\ Knupfer, M.\ Kielwein, U.\ K.\ R\"{o}ler, M.\
  S.\ Golden, J.\ Fink, F.\ M.\ F.\ de Groot, T.\ Ito, K.\ Oka, and G.\
  Kaindl, Eur.\ Phys.\ J.\ B \textbf{26}, 449 (2002).

  % Carbon nanotubes book
\bibitem{Dresselhaus_2001} Z.\ Yao, C.\ Dekker, and P.\ Avouris,
  \textit{Electrical Transport Through Single-Wall Carbon Nanotubes}, in M.\
  S.\ Dresselhaus, G.\ Dresselhaus, and P.\ Avouris, eds., \textit{Carbon
  Nanotubes:\ Synthesis, Structure, Properties and Applications} (Springer,
  2001).

  % LL param for SWCNTs http://www.springerlink.com/content/1w5vp1tjb7kgqkny/
\bibitem{Egger_1998} R.\ Egger and A.\ O.\ Gogolin, Eur.\ Phys.\ J.\ B
  \textbf{3}, 281 (1998).

  %Direct observation of Tomonaga\u2013Luttinger-liquid state in carbon
  nanotubes at low temperatures
\bibitem{Ishii_2003} H.\ Ishii, H.\ Kataura, H.\ Shiozawa, H.\ Yoshioka, H.\
  Otsubo, Y.\ Takayama, T.\ Miyahara, S.\ Suzuki, Y.\ Achiba, M.\ Nakatake,
  T.\ Narimura, M.\ Higashiguchi, K.\ Shimada, H.\ Namatame, and M.\
  Taniguchi, Nature \textbf{426}, 540 (2003).

  % LL behavior in nanotube networks
\bibitem{Shiraishi_2003} M.\ Shiraishi and M.\ Ata, Sol.\ State Commun.\
  \textbf{127}, 215 (2003).

  % LL in fractional quantum hall
\bibitem{Milliken_1996} F.\ P.\ Milliken, C.\ P.\ Umbach, and R.\ A.\ Webb,
  Sol.\ State Commun.\ \textbf{97}, 309 (1996).

  % Chiral Luttinger liquids at fractional quantum hall edges
\bibitem{Chang_2003} A.\ M.\ Chang, Rev.\ Mod.\ Phys.\ \textbf{75}, 1449
  (2003).

  % How universal is the fractional-quantum-Hall edge Luttinger liquid?
\bibitem{Mandal_2001} S.\ S.\ Mandal and J.\ K.\ Jain, Sol.\ State Commun.\
  \textbf{118}, 503 (2001).

  % Tunneling conductance between 1D wires (semiconductor heterostructures).
\bibitem{Steinberg_2006} H.\ Steinberg, O.\ M.\ Auslaender, A.\ Yacoby, J.\
  Qian, G.\ A.\ Fiete, Y.\ Tserkovnyak, B.\ I.\ Halperin, K.\ W.\ Baldwin, L.\
  N.\ Pfeiffer, and K.\ W.\ West, Phys.\ Rev.\ B \textbf{73}, 113307 (2006).

  % InSb nanowires
\bibitem{Zaitsev_2000} S.\ V.\ Zaitsev-Zotov, Y.\ A.\ Kumzerov, Y.\ A.\
  Firsov, and P.\ Monceau, J.\ Phys.:\ Condens.\ Matter \textbf{12}, L303
  (2000).

  % In2)3 nanowires
\bibitem{Liu_2005} F.\ Liu, M.\ Bao, K.\ L.\ Wang, C.\ Li, B.\ Lei, and C.\
  Zhou, Appl.\ Phys.\ Lett.\ \textbf{86}, 213101 (2005).

\bibitem{Goni_1991} A.\ R.\ Go\~ni, A.\ Pinczuk, J.\ S.\ Weiner, J.\
  M.\ Calleja, B.\ S.\ Dennis, L.\ N.\ Pfeiffer, and K.\ W.\ West,
  Phys.\ Rev.\ Lett.\ \textbf{67}, 3298 (1991).

\bibitem{Auslaender_2000} ,O.\ M.\ Auslaender, A.\ Yacoby, R.\
  dePicciotto, K.\ W.\ Baldwin, L.\ N.\ Pfeiffer, and K.\ W.\ West,
  Phys.\ Rev.\ Lett.\ \textbf{84}, 1764 (2000).

  % Ultracold atomic gases in harmonic traps (highly confined in 2 dims)
\bibitem{Moritz_2005} H.\ Moritz, T.\ Stoferle, K.\ Guenter, M.\ Kohl,
  T.\ Esslinger, Phys.\ Rev.\ Lett.\ \textbf{94}, 210401 (2005).

  % 1D Fermionic atoms
\bibitem{Recati_2003} A.\ Recati, P.\ O.\ Fedichev, W.\ Zwerger, and P.\
  Zoller, J.\ Opt.\ B:\ Quantum Semiclass.\ Opt.\ \textbf{5} S55 (2003).

  % Trapped one-dimensional Bose gas as a Luttinger liquid
\bibitem{Monien_1998} H.\ Monien, M.\ Linn, and N.\ Elstner, Phys.\ Rev.\ A
  \textbf{58}, R3395 (1998).

  % Rows of gold atoms giving a close to ideal 1D Fermi gas
\bibitem{Schaefer_2008} J.\ Sch\"{a}fer, C.\ Blumenstein, S.\ Meyer, M.\
  Wisniewski, and R.\ Claessen, Phys.\ Rev.\ Lett.\ \textbf{101}, 236802
  (2008).

\bibitem{Schulz_1993} H.\ J.\ Schulz, Phys.\ Rev.\ Lett.\ \textbf{71}, 1864
  (1993).

\bibitem{Astrakharchik_2010} G.\ E.\ Astrakharchik and M.\ D.\ Girardeau,
  arXiv:1101.0103v1 [cond-mat.quant-gas], (2010).

\bibitem{Fogler_2005} M.\ M.\ Fogler, Phys.\ Rev.\ Lett.\ \textbf{94}, 056405
  (2005).

\bibitem{Creffield_2001} C.\ E.\ Creffield, W.\ H\"{a}user, and A.\ H.\
  MacDonald, Europhys.\ Lett.\ \textbf{53}, 221 (2001).

\bibitem{Fabrizio_1994} M.\ Fabrizio, A.\ O.\ Gogolin, and S.\ Scheidl, Phys.\
  Rev.\ Lett.\ \textbf{72}, 2235 (1994).

\bibitem{Casula_2006} M.\ Casula, S.\ Sorella, and G.\ Senatore, Phys.\ Rev.\
  B \textbf{74}, 245427 (2006).

\bibitem{Shulenburger_2008} L.\ Shulenburger, M.\ Casula, G.\ Senatore, and
  R.\ M.\ Martin, Phys.\ Rev.\ B \textbf{78}, 165303 (2008).

\bibitem{Malatesta_2000} A.\ Malatesta and G.\ Senatore, J.\ Phys.\ IV
  \textbf{10}, 5 (2000).

\bibitem{Friesen_1980} W.\ I.\ Friesen and B.\ Bergersen, J.\ Phys.\ C: Solid
  St.\ Phys.\ \textbf{13}, 6627 (1980).

\bibitem{Camels_1997} L.\ Calmels, A.\ Gold, Phys.\ Rev.\ B
  \textbf{56}, 1762 (1997).

\bibitem{Tas_2003} M.\ Ta\ifmmode \mbox{\c{s}}\else \c{s}\fi{}, and
  M.\ Tomak, Phys.\ Rev.\ B \textbf{67} 235314 (2003).

\bibitem{Garg_2008} V.\ Garg, R.\ K.\ Moudgil, K.\ Kumar, and P.\ K.\
  Ahluwalia, Phys.\ Rev.\ B \textbf{78}, 045406 (2008).

\bibitem{Asgari_2007} R.\ Asgari, Solid State Commun.\ \textbf{141}, 563
  (2007).

\bibitem{Saunders_1994} V.\ R.\ Saunders, C.\ Freyria-Fava, R.\ Dovesi, and
  C.\ Roetti, Comp.\ Phys.\ Commun.\ \textbf{84}, 156 (1994).

\bibitem{Kato_1957} T.\ Kato, Comm.\ Pure Appl.\ Math.\ \textbf{10}, 151
  (1957).

\bibitem{Lieb_1962} E.\ Lieb and D.\ Mattis, Phys.\ Rev.\ \textbf{125}, 164
  (1962).

\bibitem{Malatesta_thesis} A.\ Malatesta, \textit{Quantum Monte Carlo study of
    a model one-dimensional electron gas}, PhD Thesis, University of Trieste,
    Trieste, 1999.

\bibitem{casino2} R.\ J.\ Needs, M.\ D.\ Towler, N.\ D.\ Drummond, and P.\
  L\'opez R\'{\i}os, J.\ Phys.:\ Condens.\ Matter \textbf{22}, 023201 (2010).

%\bibitem{Wigner_1934} E.\ P.\ Wigner, Phys.\ Rev.\ \textbf{46}, 1002
%  (1934).

%\bibitem{Drummond_2004} N.\ D.\ Drummond, Z.\ Radnai, J.\ R.\ Trail,
%  M.\ D.\ Towler, and R.\ J.\ Needs, Phys.\ Rev.\ B \textbf{69}, 085116
%  (2004).

%  \bibitem{Tserkovnyak_2003} Y.\ Tserkovnyak, B.\ Halperin, O.\ Auslaender, and
%  A.\ Yacoby, Phys.\ Rev.\ B \textbf{68}, 125312 (2003).

%  \bibitem{Hirsch_1982} J.\ E.\ Hirsch, R.\ L.\ Sugar, D.\ J.\ Scalapino,
% and R.\ Blankenbecler, Phys.\ Rev.\ B \textbf{26}, 5033 (1982).

%  \bibitem{Hu_1993} B.\ Yu-Kuang Hu and S.\ Das Sarma, Phys.\ Rev.\ B
%  \textbf{48}, 5469 (1993).

%  \bibitem{Agosti_1998} D.\ Agosti, F.\ Pederiva, E.\ Lipparini, and K.\
%  Takayanagi, Phys.\ Rev.\ B \textbf{57}, 14869 (1998).

\bibitem{Foulkes_2001} W.\ M.\ C.\ Foulkes, L.\ Mitas, R.\ J.\ Needs, and G.\
  Rajagopal, Rev.\ Mod.\ Phys.\ \textbf{73}, 33 (2001).

\bibitem{Umrigar_1988a} C.\ J.\ Umrigar, K.\ G.\ Wilson, and J.\ W.\ Wilkins,
  Phys.\ Rev.\ Lett.\ \textbf{60}, 1719 (1988).

\bibitem{Kent_1999} P.\ R.\ C.\ Kent, R.\ J.\ Needs, and G.\ Rajagopal, Phys.\
  Rev.\ B \textbf{59}, 12344 (1999).

\bibitem{Ndd_newopt} N.\ D.\ Drummond and R.\ J.\ Needs, Phys.\ Rev.\ B
  \textbf{72}, 085124 (2005).

\bibitem{Umrigar_emin} C.\ J.\ Umrigar, J.\ Toulouse, C.\ Filippi, S.\
  Sorella, and R.\ G.\ Hennig, Phys.\ Rev.\ Lett.\ \textbf{98}, 110201 (2007).

\bibitem{Ceperley_1980} D.\ M.\ Ceperley and B.\ J.\ Alder, Phys.\ Rev.\
  Lett.\ \textbf{45}, 566 (1980).

\bibitem{ndd_jastrow} N.\ D.\ Drummond, M.\ D.\ Towler, and R.\ J.\ Needs,
  Phys.\ Rev.\ B \textbf{70}, 235119 (2004).

\bibitem{Pablo_2006} P.\ L\'{o}pez R\'{i}os, A.\ Ma, N.\ D.\ Drummond, M.\ D.\
  Towler, and R.\ J.\ Needs, Phys.\ Rev.\ E \textbf{74}, 066701 (2006).

\bibitem{Chiesa_2006} S.\ Chiesa, D.\ M.\ Ceperley, R.\ M.\ Martin, and M.\
  Holzmann, Phys.\ Rev.\ Lett.\ \textbf{97}, 076404 (2006).

\bibitem{Rajagopal_1995} G.\ Rajagopal, R.\ J.\ Needs, A.\ James, S.\ D.\
  Kenny, and W.\ M.\ C.\ Foulkes, Phys.\ Rev.\ B \textbf{51}, 10591
  (1995).

\bibitem{Ndd_2009} N.\ D.\ Drummond and R.\ J.\ Needs, Phys.\ Rev.\ Lett.\
  \textbf{102}, 126402 (2009).

\bibitem{Holzmann_2009} M.\ Holzmann, B.\ Bernu, V.\ Olevano, R.\ M.\ Martin,
  and D.\ M.\ Ceperley, Phys.\ Rev.\ B \textbf{79}, 041308(R) (2009).

\bibitem{Twist_averaging} C.\ Lin, F.\ H.\ Zong, and D.\ M.\ Ceperley, Phys.\
  Rev.\ E \textbf{64}, 016702 (2001).

\bibitem{Mattis_1965} D.\ C.\ Mattis and E.\ H.\ Lieb, J.\ Math.\ Phys.\
  \textbf{6}, 304 (1965).

\bibitem{Schulz_1990} H.\ J.\ Schulz, Phys.\ Rev.\ Lett.\ \textbf{64}, 2831
(1990).

%\bibitem{Monkhorst_1976} H.\ J.\ Monkhorst and J.\ D.\ Pack, Phys.\ Rev.\
%  B \textbf{13}, 5188 (1976).

%\bibitem{Martin_elecstruc} R.\ M.\ Martin, \textit{Electronic
%    Structure}, (Cambridge University Press, Cambridge, 2004).

\bibitem{Wang_2001} D.\ W.\ Wang, A.\ J.\ Millis, and S.\ Das Sarma, Phys.\
  Rev.\ B \textbf{64}, 193307 (2001).

\bibitem{Popovic_2001} Z.\ V.\ Popovic, V.\ A.\ Ivanov, O.\ P.\ Khoung, and
  V.\ V.\ Moshchalkov, Synth.\ Metals \textbf{124}, 421 (2001).

\bibitem{Vescoli_1998} V.\ Vescoli, L.\ Degiorgi, W.\ Henderson, G.\
  Gr\"{u}ner, K.\ P.\ Starkey, and L.\ K.\ Montgomery, Science \textbf{281},
  1181 (1998).

\bibitem{Georges_2000} A.\ Georges, T.\ Giamarchi, and N.\ Sandler, Phys.\
  Rev.\ B \textbf{61}, 16393 (2000).

\end{thebibliography}
\end{document}